\documentclass[12pt]{article}

\usepackage{hyperref}
\hypersetup{
    colorlinks=true,
    linkcolor=blue,
    citecolor=blue,
    filecolor=magenta,
    urlcolor=green,
}
\usepackage{bm}
\usepackage{natbib}
\usepackage{graphicx}
\usepackage{multirow}
\usepackage{enumitem}
\usepackage{amsmath,amsthm,amssymb}
\usepackage{chngcntr}
\usepackage{mathrsfs}
\usepackage{authblk}
\usepackage{makecell}
\usepackage{subcaption}
\usepackage{float}
\usepackage[linesnumbered,ruled,vlined]{algorithm2e}

\def\f{Fr\'echet }
\DeclareMathOperator*{\argmin}{arg\,min}
\DeclareMathOperator{\Var}{Var}

\DeclareMathOperator{\E}{E}

\newcommand{\single}{\spacingset{1}}

\newtheorem{proposition}{Proposition}[section]
\newtheorem{theorem}{Theorem}[section]
\newtheorem{corollary}{Corollary}[section]
\newtheorem{assumption}{Assumption}[section]

\newtheorem{remark}{Remark}[section]
\newtheorem{example}{Example}

\begin{document}

\def\spacingset#1{\renewcommand{\baselinestretch}%
{#1}\small\normalsize} \spacingset{1}

\title{\bf Geodesic Synthetic Control Methods for Random Objects and Functional Data}
\author[1,*]{Daisuke Kurisu}
\author[2,*]{Yidong Zhou}
\author[3]{Taisuke Otsu}
\author[2,$^{\dagger}$]{Hans-Georg M\"uller}
\affil[1]{Faculty of Economics, The University of Tokyo, Japan}
\affil[2]{Department of Statistics, University of California, Davis, USA}
\affil[3]{Department of Economics, London School of Economics, UK}
\maketitle
\renewcommand*{\thefootnote}{\fnsymbol{footnote}}
\footnotetext[1]{The first two authors contributed equally to this work and are listed alphabetically.}
\footnotetext[2]{Corresponding author: hgmueller@ucdavis.edu.}
\footnotetext[3]{D.K. was supported in part by JSPS KAKENHI Grant Numbers 23K12456, 25K00624, and 26K00323. H.G.M. was partially supported by NSF grant DMS-2310450.}

\bigskip
\begin{abstract}
We introduce a geodesic synthetic control method for causal inference with panel outcomes that are random objects in a geodesic metric space. Examples include distributions, compositions, networks, symmetric positive-definite matrices, trees and functional data, among other data types that require a geometry beyond Euclidean vector spaces. The proposed method replaces Euclidean weighted averages by weighted Fr\'echet means and defines treatment effects as geodesics connecting untreated and treated potential outcomes. We develop a causal model with stochastic perturbations for object-valued untreated outcomes, establish consistency of the estimated weights for the perturbed population target, derive a bound on post-treatment prediction error, and give sufficient conditions for consistent recovery of the untreated counterfactual. For uncertainty quantification, we propose an intrinsic metric conformal calibration procedure that yields prediction regions for untreated counterfactual objects, confidence sets for geodesic treatment effects, confidence intervals for their magnitudes, and global no-effect tests. The method is illustrated through simulation studies for networks and symmetric positive-definite matrices, and through applications to employment composition changes following the 2011 Great East Japan Earthquake and the impact of abortion liberalization on fertility patterns in East Germany.
\end{abstract}

\noindent%
\vfill

\newpage
\newcommand{\double}{\spacingset{1.75}}
\double 

\section{Introduction}
Since the seminal work of \citet{AbGa03} and \citet{AbDiHa10}, synthetic control methods (SCMs) have become a widely used framework for evaluating causal effects of policy interventions in observational panel and longitudinal studies. SCMs are designed for settings in which some units are exposed to an intervention while others are not, with outcomes observed both before and after the intervention. The central idea is to construct, from the untreated units, a weighted synthetic control that closely reproduces the pre-treatment trajectory of the treated unit, and then to use this synthetic control as a counterfactual benchmark in the post-treatment period. The method has been applied in a wide range of comparative case studies; see \citet{abad:21} for a recent survey.

Substantial methodological developments have extended SCMs in several directions. These include penalized synthetic controls that improve matching and regularization \citep{AbLh21}, SCMs for high-dimensional settings \citep{RoSaKi17}, connections with matrix completion methods \citep{AmShSh18}, and inferential procedures based on conformal prediction and robust testing \citep{ChWuZh21a,ChWuZh21b}. Bias-corrected and augmented synthetic control estimators have been developed to address imperfect pre-treatment fit \citep{BeFeRo21}, with further extensions to staggered adoption settings \citep{ben:22}. In addition, \citet{ArAtHiImWa21} introduced synthetic difference-in-differences, which combines features of SCM and difference-in-differences. These developments, however, have primarily focused on Euclidean outcomes.

Many modern applications involve outcomes that are not naturally represented as scalars or vectors. Examples include distributions \citep{mull:16:1,pana:20,pete:22}, compositional data \citep{scea:11}, manifold-valued data \citep{patr:15,kuri:26}, networks \citep{mull:22:11,seve:22}, functional data \citep{hsin:15,mull:16:3}, and tree-structured data \citep{bill:01}. These objects are often best viewed as elements of a metric space endowed with a geometry appropriate for the type of data one observes. Common choices include the Wasserstein metric for distributions, the arc-length metric for compositional data, Riemannian metrics for manifold-valued data, the Frobenius or power metric when representing networks through their graph Laplacians, the $L^2$ metric for functional data, and the BHV metric for tree spaces. Many such spaces are geodesic metric spaces, but they generally lack linear operations, such as addition, subtraction, and scalar multiplication, on which classical SCMs rely. This creates a fundamental obstacle for causal inference, since both the construction of a weighted synthetic control and the definition of a treatment effect must be reformulated intrinsically in the geometry of the outcome space. General approaches to address related challenges for data analysis in metric spaces are discussed in \citet{mull:24}.

Causal inference for non-Euclidean and geodesic outcomes is a recent and rapidly developing area. Existing contributions include methods for average treatment effects \citep{lin:23}, difference-in-differences \citep{toro:24}, synthetic controls \citep{guns:23,GuHsLe24}, and regression discontinuity designs \citep{va:25}, with many of these focusing on distributional outcomes. For more general random objects in geodesic metric spaces, \citet{kuri:24} developed a framework for geodesic average treatment effects, and \citet{zhou:25} proposed geodesic difference-in-differences. These works demonstrate the promise of causal methods for object-valued outcomes, but a general synthetic control framework for a broad class of geodesic metric spaces along with consistency theory and intrinsic inference has been missing.

In this paper we propose such a framework, the geodesic synthetic control method (GSC),  for causal inference with outcomes taking values in a geodesic metric space. The proposed method replaces Euclidean weighted averages by weighted Fr\'echet means and defines treatment effects through geodesics connecting untreated and treated potential outcomes. This provides an intrinsic extension of the classical SCM to outcomes such as distributions, compositions, networks, SPD matrices, and functions. For distributional outcomes equipped with the Wasserstein metric the proposed framework includes the distributional synthetic control approach of \citet{guns:23} as a special case, and encompasses a much wider class of geodesic metrics for distributions such as the Fisher--Rao metric
; importantly, it extends beyond distributional data to a much wider class of random objects.

For outcomes in general geodesic metric spaces, we develop statistical theory and inference for GSC under a perturbed geodesic causal model, where untreated object-valued outcomes are generated by random perturbation maps acting on latent unit-specific objects. This formulation accommodates idiosyncratic time- and unit-specific variation and reduces, in the Euclidean case, to a latent-factor-type model with additive noise \citep{AbDiHa10}. We establish consistency of the estimated GSC weights for the minimizer set of the limiting perturbed criterion, derive a post-treatment prediction-error bound, and give sufficient conditions for consistent recovery of the untreated counterfactual. For uncertainty quantification, we propose an intrinsic inference procedure based on metric conformal calibration, using pre-treatment prediction errors measured directly in the outcome metric to construct prediction regions for untreated counterfactual objects, confidence sets for geodesic treatment effects, and a global no-effect test. We assess the method through simulations for networks and symmetric positive-definite matrices and illustrate it in two empirical applications: the effect of the 2011 Great East Japan Earthquake on employment composition in Miyagi Prefecture and the effect of the 1972 abortion legislation on age-specific fertility patterns in East Germany.

The paper is organized as follows. Section~\ref{sec:pre} reviews basic concepts from metric geometry and introduces the geodesic metric spaces used throughout the paper. Section~\ref{sec:met} presents the GSC estimator, the perturbed causal model and theory. Section~\ref{sec:inf} develops metric conformal calibration for inference. 
Section~\ref{sec:sim} provides implementation details and network simulations; parallel results for SPD matrices are in Section~S.5 of the Supplement. Section~\ref{sec:app} presents the empirical applications to employment composition in Japan and fertility patterns in East Germany. Section~\ref{sec:dis} discusses extensions, including augmented GSC and geodesic synthetic difference-in-differences, and concludes. Proofs and additional technical details are provided in the Supplementary Material.

\section{Preliminaries on metric geometry}\label{sec:pre}
Let $(\mathcal{M},d)$ be a separable metric space such that $(\mathcal M,\mathcal B(\mathcal M))$ is a standard Borel space, where $\mathcal B(\mathcal M)$ is the Borel $\sigma$-field generated by the metric topology. Let $(\Omega,\mathcal A,\Pr)$ be a probability space. An $\mathcal M$-valued random object is a measurable map $Y:(\Omega,\mathcal A)\to(\mathcal M,\mathcal B(\mathcal M))$. The random object $Y$ has a bounded essential range if there exist $\nu_0\in\mathcal M$ and $D<\infty$ such that $\Pr\{d(\nu_0,Y)\leq D\}=1$. A collection $\mathcal Y$ of $\mathcal M$-valued random objects has uniformly bounded essential ranges if the same $\nu_0$ and $D$ satisfy this condition for every $Y\in\mathcal Y$.  Throughout, the latent signals and random-object outcomes considered below, including those in triangular arrays, are assumed jointly to have uniformly bounded essential ranges over all unit, time, and array indices.

We briefly review the metric-geometric concepts needed to study outcomes taking values in $\mathcal{M}$. A \textit{curve} in $\mathcal{M}$ is a continuous map $\gamma : [0, 1] \to \mathcal{M}$, whose length is defined as $\ell(\gamma) = \sup \sum_{i=0}^{I-1} d(\gamma(t_i), \gamma(t_{i+1}))$, where the supremum is taken over all possible finite partitions $0 = t_0 \leq \cdots \leq t_I = 1$. A curve $\gamma$ is called a \textit{geodesic} if it satisfies $d(\gamma(s), \gamma(t))=|t-s|d(\gamma(0), \gamma(1))$ for all $s, t\in[0, 1]$. The metric space $\mathcal{M}$ is called a \textit{geodesic space} if, for any two points $\alpha, \beta \in \mathcal{M}$, there exists a geodesic connecting them, denoted $\gamma_{\alpha, \beta}$. If this geodesic is unique for every pair of points in $\mathcal{M}$, the space is referred to as a \textit{uniquely geodesic space} \citep{brids:99}. 

Examples of uniquely geodesic spaces that frequently arise in applications are as follows. 
\begin{example}[Networks]\label{exm:net}
Consider the space of simple, undirected, weighted networks with $m$ nodes. Each network can be uniquely represented by its graph Laplacian matrix. The space of graph Laplacians $\mathcal{L}$ serves as a natural framework for characterizing network structures \citep{kola:14,seve:22,mull:22:11}. When endowed with the Frobenius metric, defined as $d_F(L_1, L_2) = \|L_1 - L_2\|_F$ for $L_1, L_2 \in \mathcal{L}$, the space $\mathcal{L}$ becomes a uniquely geodesic space. In this setting, the geodesic between $L_1$ and $L_2$ is given by the linear interpolation $\gamma_{L_1, L_2}(t) = (1 - t)L_1 + tL_2$ for $t \in [0, 1]$.
\end{example}

\begin{example}[Compositional Data]\label{exm:com}
Compositional data, represented by vectors of non-negative proportions summing to one, reside in the simplex $\Delta^{d-1} = \{\bm{y} \in \mathbb{R}^d : y_j \geq 0, \sum_{j=1}^d y_j = 1\}$. Adopting the square-root transformation provides a map from $\Delta^{d-1}$ to the positive orthant of the unit sphere $\mathcal{S}_+^{d-1}$ \citep{scea:11}, where the induced geodesic distance on $\mathcal{S}_+^{d-1}$ is the arc-length distance $d_g(\bm{z}_1, \bm{z}_2) = \arccos(\bm{z}_1^\top \bm{z}_2)$. Let $\theta = \arccos(\bm{z}_1^\top \bm{z}_2)$ be the angle between $\bm{z}_1$ and $\bm{z}_2$ and $\|\cdot \|$ be the standard Euclidean norm. The geodesic connecting $\bm{z}_1$ and  $\bm{z}_2$ is then
\[\gamma_{\bm{z}_1, \bm{z}_2}(t)=\cos(t \theta) \bm{z}_1 + \sin(t \theta) \frac{\bm{z}_2 - (\bm{z}_1^\top \bm{z}_2) \bm{z}_1}{\|\bm{z}_2 - (\bm{z}_1^\top \bm{z}_2) \bm{z}_1\|},\quad t \in [0, 1].\]

\end{example}

\begin{example}[Symmetric Positive-Definite Matrices]\label{exm:cov}
The space of $m \times m$ symmetric positive-definite (SPD) matrices, denoted by $\mathrm{Sym}^+_m$, plays a crucial role in various fields such as computer vision \citep{cher:16}, signal processing \citep{arna:13} and neuroimaging \citep{dryd:09}. Various metrics have been introduced to endow $\mathrm{Sym}^+_m$ with a meaningful geometric structure. The Frobenius metric $d_F$ induces unique geodesics that are linear interpolations, $\gamma_{A, B}(t) = (1 - t)A + tB$ for $A, B \in \mathrm{Sym}^+_m$. The space $\mathrm{Sym}^+_m$ also forms a uniquely geodesic space under various other metrics such as 
the Log-Euclidean metric \citep{arsi:07}, the power metric family \citep{dryd:09}, or the Log-Cholesky metric \citep{lin:19:1}. For instance, under the Log-Euclidean metric, the unique geodesic connecting $A, B \in \mathrm{Sym}^+_m$ is $\gamma_{A, B}(t) = \exp\{(1-t)\log A+t\log B\}$, where $\exp$ and $\log$ denote the matrix exponential and matrix logarithm, respectively.  %
\end{example}

\begin{example}[Probability Distributions]\label{exm:mea}
Let $\mathcal{W}$ denote the space of univariate probability distributions on $\mathbb{R}$ with finite second moments, equipped with the 2-Wasserstein distance $d_{\mathcal{W}}(\mu, \nu) = \left(\int_0^1 \{F_{\mu}^{-1}(s) - F_{\nu}^{-1}(s)\}^2 ds \right)^{1/2}$, where $F_{\mu}^{-1}$ and $F_{\nu}^{-1}$ are the quantile functions of the probability measures $\mu$ and $\nu$, respectively. The Wasserstein space $\mathcal{W}$ is a complete and separable uniquely geodesic space \citep{ambr:08}, with geodesics  given by McCann's interpolant \citep{mcca:97}: $\gamma_{\mu, \nu}(t) = \{\mathrm{id} + t (F_{\nu}^{-1} \circ F_{\mu} - \mathrm{id})\}_\# \mu$, $t \in [0, 1]$, where $\mathrm{id}$ denotes the identity map, $F_{\mu}$ is the cumulative distribution function of $\mu$, and $\tau_\# \mu$ denotes the pushforward measure of $\mu$ by $\tau$.
In addition, several metrics that are computationally tractable for both univariate and multivariate distributions are available. One example is the Fisher--Rao metric $d_{FR}(\mu, \nu)= \arccos\left(\int_{\mathbb{R}}\sqrt{f_\mu(x)f_\nu(x)}dx\right)$, where $f_\mu$ and $f_\nu$ are the probability density functions of the measures  $\mu$ and $\nu$, respectively.  
\end{example}

\begin{example}[Functional Data]\label{exm:fun}
In functional data analysis, one considers data consisting of functions that are considered to be the realizations of a square integrable stochastic process, usually assumed to be defined on a finite interval. Such data are common in the social, physical and life sciences \citep{rams:05,hsin:15,mull:16:3}. Typically, functional data are assumed to be situated in the  Hilbert space  $L^2(\mathcal{T})$ of square integrable functions, where $\mathcal{T}$ is a compact interval, most commonly a time domain. In contrast to the other metric spaces we consider in this paper, this space is a vector space and thus permits algebraic operations. It is equipped with the standard $L^2$ metric $d_{L^2}(g_1, g_2) = \left(\int_{\mathcal{T}}\{g_1(t)-g_2(t)\}^2 dt\right)^{1/2}$, making $L^2(\mathcal{T})$ a uniquely geodesic space, with the unique geodesic between $g_1$ and $g_2$  given by $\gamma_{g_1,g_2}(t) = (1 - t)g_1 + tg_2$, $t \in [0,1]$.
\end{example}

\section{Geodesic synthetic control methods}\label{sec:met}
\subsection{Construction of the estimator}

We begin with the conventional SCM for Euclidean outcomes. Suppose outcomes $Y_{j,t}\in\mathbb{R}^q$ are observed for $J+1$ units over $T$ time periods, indexed by $j=1,\dots,J+1$ and $t=1,\dots,T$. Let $T_0>1$ denote the last pre-treatment period, and assume that only unit $j=1$ receives the intervention for $t>T_0$. For unit $j$ at time $t$, we denote potential outcomes under treatment and no treatment by $Y_{j,t}(I)$ and $Y_{j,t}(N)$, respectively. We observe $Y_{1,t}=Y_{1,t}(I)$ for $t>T_0$ and $Y_{j,t}=Y_{j,t}(N)$ otherwise. The causal effect of interest is
\[
\tau_t = Y_{1,t}(I) - Y_{1,t}(N), \quad t=T_0+1,\dots,T.
\]

In Euclidean settings, the synthetic control estimator constructs a convex combination of the control units that best reproduces the pre-treatment trajectory of the treated unit. Let $\Delta^{J-1}=\{\bm{w}=(w_2, \ldots, w_{J+1})^\top\in\mathbb{R}^J:w_j\geq0,\sum_{j=2}^{J+1}w_j=1\}$ denote the probability simplex, and $\widehat{\bm{w}}=(\widehat{w}_2,\dots,\widehat{w}_{J+1})^\top\in\Delta^{J-1}$ be the optimal weight vector. The synthetic control outcome for the treated unit is defined as
\[
\widehat Y_{1,t}^{(\mathrm{SC})} = \sum_{j=2}^{J+1}\widehat{w}_j Y_{j,t},\quad t=T_0+1,\dots,T,
\]
where the weights are chosen to minimize the mean squared pre-treatment discrepancy,
\[
\widehat{\bm{w}} \in \argmin_{\bm{w}\in\Delta^{J-1}} \frac{1}{T_0}\sum_{t=1}^{T_0}\big\|Y_{1,t}-Y_{2:J+1,t}^{(\bm{w})}\big\|^2,\quad
Y_{2:J+1,t}^{(\bm{w})}=\sum_{j=2}^{J+1}w_jY_{j,t}.
\]
The SCM estimator of the causal effect is then given by
\[
\widehat\tau_t^{(\mathrm{SC})}=Y_{1,t}-\widehat Y_{1,t}^{(\mathrm{SC})},\quad t=T_0+1,\dots,T.
\]

The construction of SCM relies on two key components: the weighted mean of the control outcomes $Y_{2:J+1,t}^{(\bm{w})}$ and the difference $Y_{1,t} - \widehat Y_{1,t}^{(\mathrm{SC})}$. Both are natural in Euclidean space but are not directly available for non-Euclidean outcomes, where algebraic operations like addition, subtraction, or scalar multiplication are not defined.

For general metric spaces, the appropriate analogue of the weighted mean is the weighted Fr\'echet mean \citep{frec:48,gine:21}, defined for objects $Y_{2,t},\dots,Y_{J+1,t}\in\mathcal{M}$ as
\[
Y_{2:J+1,t}^{(\bm{w})} = \argmin_{\nu\in\mathcal{M}} \sum_{j=2}^{J+1} w_j d^2(\nu, Y_{j,t}),
\]
which reduces to the weighted Euclidean average when $\mathcal{M}=\mathbb{R}^q$ and $d$ is the Euclidean distance. 
When the outcome space is geodesic, a suitable generalization for otherwise unavailable differences exists: in the special case of Euclidean space, the difference $\beta-\alpha$ can be equivalently interpreted as movement along the straight line (geodesic) from $\alpha$ to $\beta$, which motivates defining causal effects via geodesics \citep{kuri:24}.

Accordingly, in a uniquely geodesic space $(\mathcal{M},d)$ we define the causal effect of the intervention as the geodesic connecting the untreated and treated potential outcomes,
\[\tau_t = \gamma_{Y_{1,t}(N),Y_{1,t}(I)}, \quad t=T_0+1,\dots,T.\]
The uniqueness of geodesics ensures that $\tau_t$ is well defined. This definition preserves the interpretation of $\tau_t$ as encoding both the size  and direction of change.

\begin{remark}
    As pointed out by a referee, one may alternatively represent the causal effect through the ordered pair of potential outcomes $\left(Y_{1,t}(N),Y_{1,t}(I)\right)$ rather than the geodesic $\tau_t=\gamma_{Y_{1,t}(N),Y_{1,t}(I)}$. This ordered-pair representation contains the same endpoint information and would allow a more general formulation without requiring unique geodesics. We adopt the geodesic formulation because it provides a direct intrinsic analogue of the classical Euclidean treatment effect $Y_{1,t}(I)-Y_{1,t}(N)$: in Euclidean space, subtraction represents a displacement from the untreated outcome to the treated outcome, while in a uniquely geodesic metric space the geodesic represents the corresponding intrinsic movement, with its length giving the effect size  and its path describing the direction of change. Moreover, the geodesic structure is also used in the causal model in Section~\ref{subsec:cm}, where geodesic interpolation between a time-specific reference object and a unit-specific latent object provides an analogue of the latent-factor structure in classical synthetic control \citep{AbDiHa10}. The uniquely geodesic assumption is not overly restrictive for our purposes, as many spaces of practical interest are uniquely geodesic, including the spaces considered in Examples~\ref{exm:net}--\ref{exm:fun} as well as phylogenetic tree spaces equipped with the BHV metric \citep{bill:01}.
\end{remark}

To estimate $\tau_t$, we extend the weight construction from SCM using weighted Fr\'echet means. The synthetic control outcome for the treated unit is
\[
\widehat Y_{1,t}^{(\mathrm{GSC})} = \argmin_{\nu\in\mathcal{M}} \sum_{j=2}^{J+1} \widehat{w}_j d^2(\nu,Y_{j,t}),
\]
with weights chosen as
\begin{equation}\label{eq:w-t}
\widehat{\bm{w}} \in \argmin_{\bm{w}\in\Delta^{J-1}} \frac{1}{T_0}\sum_{t=1}^{T_0} d^2\!\left(Y_{1,t}, Y_{2:J+1,t}^{(\bm{w})}\right),\quad Y_{2:J+1,t}^{(\bm{w})} = \argmin_{\nu\in\mathcal{M}} \sum_{j=2}^{J+1} w_j d^2(\nu, Y_{j,t}).
\end{equation}
The \emph{geodesic synthetic control} (GSC) estimator of the treatment effect is then
\[
\widehat \tau_t^{(\mathrm{GSC})} = \gamma_{\widehat Y_{1,t}^{(\mathrm{GSC})}, Y_{1,t}}, \quad t=T_0+1,\dots,T.
\]

\subsection{Causal model for the geodesic synthetic control method}\label{subsec:cm}
To analyze the statistical properties of the GSC estimator, we introduce a causal model for the untreated potential outcomes that allows for stochastic perturbations. Each unit $j$ is associated with a unit-specific latent object $U_j\in\mathcal{M}$, and the untreated outcome at time $t$ is generated as
\begin{equation}\label{eq:GSC-model}
Y_{j,t}(N)=g_t(\mathcal{P}_{j,t}(U_j)),
\quad j=1,\ldots,J+1,\quad t=1,\ldots,T,
\end{equation}
where $g_t:\mathcal{M}\to\mathcal{M}$ is a deterministic measurable map and $\mathcal{P}_{j,t}:\mathcal{M}\to\mathcal{M}$ is a random perturbation map. The latent object $U_j$ captures persistent unit-level heterogeneity, while $\mathcal{P}_{j,t}$ represents time- and unit-specific stochastic deviations around this latent structure, allowing  the observed untreated outcomes to deviate from the latent geodesic signal. Throughout the asymptotic analysis, the latent objects $U_1,\ldots,U_{J+1}$ are treated as fixed, and all probabilities and expectations are taken with respect to the perturbation maps. 

For illustration, consider a geodesic interpolation,
\begin{equation}\label{eq:GSC-model-g}
g_t(\mathcal{P}_{j,t}(U_j))=\gamma_{\mu_t,\mathcal{P}_{j,t}(U_j)}(\alpha_t),\quad \alpha_t\in(0,1),
\end{equation}
where $\mu_t\in\mathcal{M}$ is a time-specific reference object. In this model, the untreated outcome is obtained by moving from the common time-specific object $\mu_t$ toward the perturbed unit-specific object $\mathcal{P}_{j,t}(U_j)$ along the geodesic connecting them. For Euclidean outcomes with the Euclidean metric, taking an additive perturbation $\mathcal{P}_{j,t}(U_j)=U_j+\varepsilon_{j,t}$ yields
\[
Y_{j,t}(N)=(1-\alpha_t)\mu_t+\alpha_t U_j+\alpha_t\varepsilon_{j,t},
\]
which corresponds to a latent-factor-type model with additive idiosyncratic noise. 

\begin{figure}[tb]
    \single
    \centering
    \includegraphics[width=0.8\linewidth]{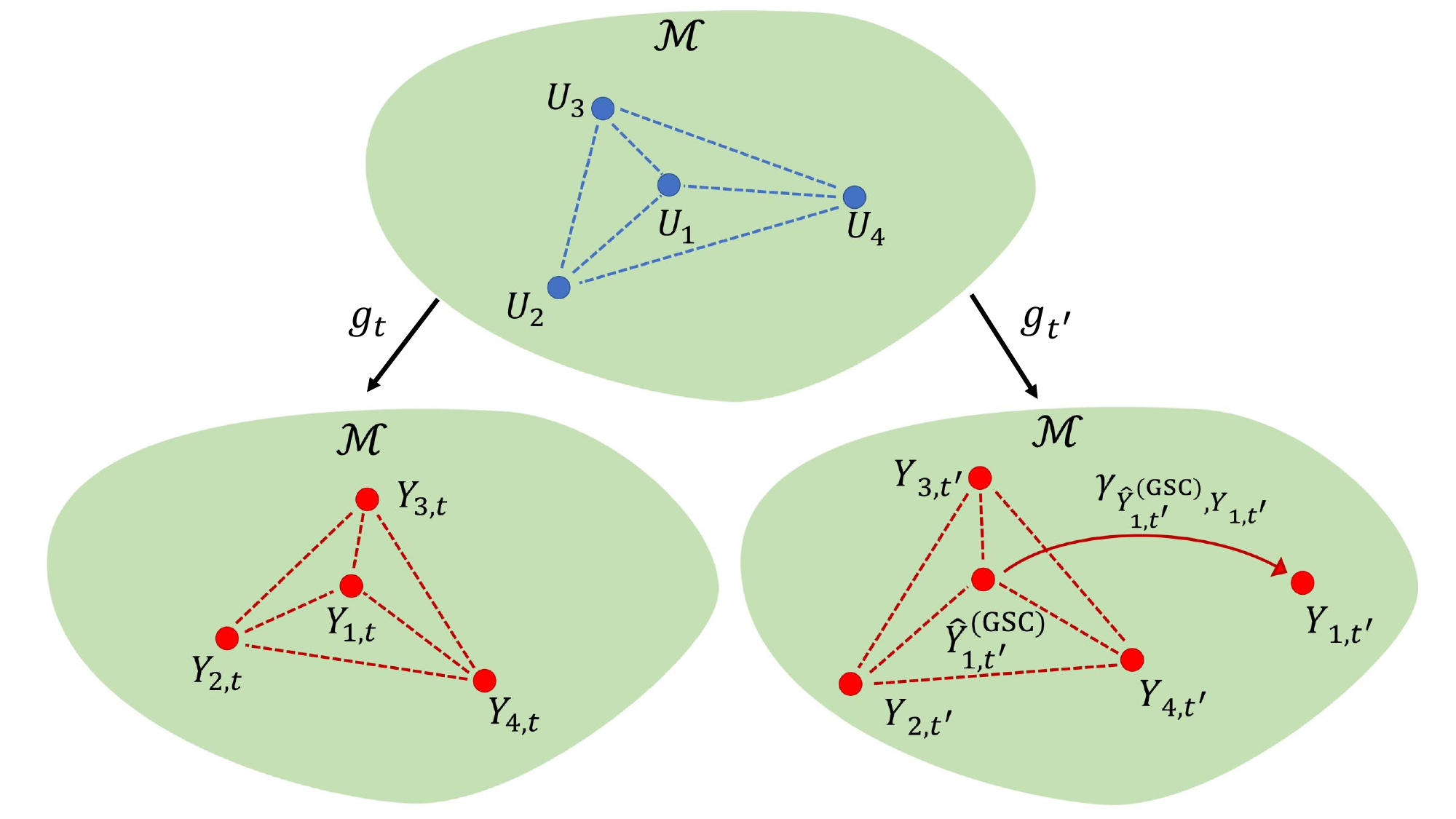}
    \caption{Causal framework for the geodesic synthetic control method. Circles represent random objects in the metric space $\mathcal{M}$. The top panel shows latent unit objects; under model~\eqref{eq:GSC-model}, random perturbation maps act on these objects before $g_t$ and are omitted from the diagram for visual clarity. For $t\leq T_0$, the treated outcome $Y_{1,t}$ is matched by a weighted Fr\'echet mean of controls. For $t'>T_0$, the estimator compares $\widehat{Y}_{1,t'}^{(\mathrm{GSC})}$ with the observed $Y_{1,t'}$, and the connecting geodesic represents the estimated causal effect.}
    \label{fig:GSC}
\end{figure}

Figure~\ref{fig:GSC} summarizes the causal framework. Its top panel shows the latent unit objects, while the lower panels show the observed pre- and post-treatment outcomes used to construct the GSC counterfactual and geodesic effect. Under model~\eqref{eq:GSC-model}, the perturbation maps act on the latent objects before the time-specific map $g_t$; they are omitted from the diagram for visual clarity.

The formulation in \eqref{eq:GSC-model} places the latent unit characteristic directly in the outcome space $\mathcal{M}$. This can be relaxed by allowing the latent characteristic to originate in another metric space $\mathcal{M}'$ and enter the outcome space through a time-specific map. Specifically, one may consider $Y_{j,t}(N)=g_t(\mathcal{P}_{j,t}(U_{j,t}))$, where $U_{j,t}\in\mathcal{M}$ may be represented as $U_{j,t}=\Phi_t(V_j)$ for some $V_j\in\mathcal{M}'$ and measurable map $\Phi_t:\mathcal{M}'\to\mathcal{M}$. In the Euclidean scalar case, taking $\mathcal{M}'=\mathbb{R}^d$ and $\Phi_t(v)=\nu_t^\top v$ for some $\nu_t \in \mathbb{R}^d$ recovers the usual linear factor representation. 

We next distinguish between two weighted Fr\'echet means. For a weight vector
$\bm{w}\in\Delta^{J-1}$, define the perturbed synthetic outcome
\[\widehat{g}_{2:J+1,t}^{(\bm{w})}:= \argmin_{\nu \in \mathcal{M}}\sum_{j=2}^{J+1}w_jd^2(\nu, g_t(\mathcal{P}_{j,t}(U_j))),\]
and the corresponding latent synthetic outcome
\[g_{2:J+1,t}^{(\bm{w})}:= \argmin_{\nu\in\mathcal{M}} \sum_{j=2}^{J+1} w_j d^2(\nu, g_t(U_j)).\]
The first quantity is the synthetic control constructed from the observed, perturbed control outcomes, used to estimate the GSC weights; the second is its latent-signal counterpart that is used in the identification condition.

For each time point $t$, let $\mathbb{W}_t^* = \left\{ \bm{w}\in\Delta^{J-1} : d(g_t(U_1), g_{2:J+1,t}^{(\bm{w})})=0 \right\}$ denote the set of weights that exactly reconstruct the latent signal of the treated unit at time $t$. The following assumption ensures that the synthetic objects are well defined and formalizes exact reconstruction of the latent signal. 

\begin{assumption}\label{ass:GSC1}
\hfill
\begin{itemize}
\item[(i)] For each $\bm{w}\in\Delta^{J-1}$ and $t=1, \ldots, T$, the weighted Fr\'echet means $g_{2:J+1,t}^{(\bm{w})}$ and $\widehat{g}_{2:J+1,t}^{(\bm{w})}$ exist uniquely.
\item[(ii)] There exists a nonempty set $\mathcal{W}_0\subset\Delta^{J-1}$, independent of $T_0$, such that $\mathcal{W}_0\subset\bigcap_{t=1}^{T_0}\mathbb{W}_t^*$ for all sufficiently large $T_0$. Moreover, for every fixed $h\geq1$, any pre-treatment latent balancing weight also reconstructs the treated unit's latent signal at $t=T_0+h$, that is, $\bigcap_{t=1}^{T_0}\mathbb{W}_t^*\subset\mathbb{W}_{T_0+h}^*$ for all sufficiently large $T_0$.
\end{itemize}
\end{assumption}

Assumption~\ref{ass:GSC1}(i) ensures existence and uniqueness of the weighted Fr\'echet means; these properties hold broadly in Hadamard spaces \citep{stur:03}. Assumption~\ref{ass:GSC1}(ii) requires a common set of weights to reconstruct the treated unit's latent trajectory before treatment and at each fixed post-treatment horizon. Section~S.2 of the Supplement gives sufficient conditions for this reconstruction under the signal models and metrics treated there. 

We next consider estimation of the weights from perturbed pre-treatment outcomes. Throughout the asymptotic analysis, $J$ is fixed and $T_0\to\infty$, and post-treatment conclusions concern $t=T_0+h$ for fixed $h\geq1$. The conclusions hold jointly over a fixed finite post-treatment horizon when the stability condition below holds uniformly over that horizon.

The GSC weights are estimated from the perturbed pre-treatment outcomes by
\[
\widehat{\bm{w}} \in \argmin_{\bm{w} \in \Delta^{J-1}}\widehat{Q}(\bm{w}),\quad \widehat{Q}(\bm{w})= \frac{1}{T_0}\sum_{t=1}^{T_0}d^2(Y_{1,t}, Y_{2:J+1,t}^{(\bm{w})}),
\]
where $Y_{2:J+1,t}^{(\bm{w})}$ denotes the weighted Fr\'echet mean of the observed control outcomes at time $t$. Under \eqref{eq:GSC-model}, $Y_{2:J+1,t}^{(\bm{w})}=\widehat{g}_{2:J+1,t}^{(\bm{w})}$ for $t\leq T_0$. To describe the population target of this empirical criterion, let
\begin{equation}\label{eq:qtilde}
\widetilde{Q}(\bm{w})=\lim_{T_0 \to \infty}\E[\widehat{Q}(\bm{w})],\qquad
\widetilde{\mathcal W}=\argmin_{\bm{w} \in \Delta^{J-1}}\widetilde{Q}(\bm{w}),
\end{equation}
where $\widetilde{\mathcal W}$ is the possibly non-singleton minimizer set of the limiting perturbed criterion. For a nonempty set $A\subset\Delta^{J-1}$, write $d_1(\bm w,A)=\inf_{\bm v\in A}\|\bm w-\bm v\|_1$. 

For $x\in\mathbb{R}^p$, let $\|x\|_1=\sum_{j=1}^p |x_j|$ denote the $\ell_1$ norm. Under the standing bounded-range condition, the following assumptions ensure uniform convergence of the empirical perturbed criterion and consistency of the estimated weights for the perturbed population target.

\begin{assumption}\label{ass:GSC-weight-lim}\quad
\begin{itemize}
\item[(i)] The limiting criterion $\widetilde Q$ in \eqref{eq:qtilde} exists as a finite continuous function, $\widetilde{\mathcal W}$ is nonempty, and, for every $\varepsilon>0$,
\[
\inf_{\{\bm w\in\Delta^{J-1}:\,d_1(\bm w,\widetilde{\mathcal W})\geq\varepsilon\}}\widetilde Q(\bm w)
>\inf_{\bm w\in\Delta^{J-1}}\widetilde Q(\bm w).
\]
A measurable minimizer $\widehat{\bm w}$ of $\widehat Q$ exists and, for every $\bm w\in\Delta^{J-1}$, $\widehat Q(\bm w)\stackrel{p}{\longrightarrow}\widetilde Q(\bm w)$ as $T_0\to\infty$.
\item[(ii)] For every $\varepsilon>0$,
\[
\lim_{\delta\downarrow0}\limsup_{T_0\to\infty}
\Pr\left\{
\sup_{\substack{\bm w_1,\bm w_2\in\Delta^{J-1}\\
\|\bm w_1-\bm w_2\|_1<\delta}}
\frac{1}{T_0}\sum_{t=1}^{T_0}
d\left(\widehat g_{2:J+1,t}^{(\bm w_1)},
\widehat g_{2:J+1,t}^{(\bm w_2)}\right)>\varepsilon
\right\}=0.
\]

\end{itemize}
\end{assumption}

Assumption~\ref{ass:GSC-weight-lim}(i) provides pointwise convergence and separation of the limiting criterion, while part~(ii) imposes uniform local stability of the weighted Fr\'echet mean map. Together with the standing bounded-range condition in Section~\ref{sec:pre}, these conditions yield the stochastic equicontinuity and uniform convergence of $\widehat Q$ used to establish consistency of the estimated weights for $\widetilde{\mathcal W}$; see \citet{well:96} for the underlying empirical-process and M-estimation arguments. Section~S.3 of the Supplement gives assumption-by-assumption sufficient conditions for Assumption~\ref{ass:GSC-weight-lim} in the outcome spaces considered here. In particular, stationarity and ergodicity, or suitable weak-dependence conditions, provide a law of large numbers for the criterion in part~(i), while Hilbert-coordinate or geodesic-convexity arguments establish part~(ii).

\begin{theorem}\label{thm:GSC-validity}
Suppose that Assumption~\ref{ass:GSC1}(i) and Assumption~\ref{ass:GSC-weight-lim} hold. Then
\begin{itemize}
\item[(i)] As $T_0\to\infty$,
\[
d_1(\widehat{\bm w},\widetilde{\mathcal W})=o_p(1).
\]
Since $\widetilde{\mathcal W}$ is nonempty and compact, choose a measurable nearest-point selection $\widetilde{\bm w}\in\widetilde{\mathcal W}$ satisfying
$\|\widehat{\bm w}-\widetilde{\bm w}\|_1=d_1(\widehat{\bm w},\widetilde{\mathcal W})$.
For a fixed $h\geq1$, suppose, in addition, that the post-treatment stability condition
\[
\sup_{\substack{\bm{w}_1,\bm{w}_2 \in \Delta^{J-1}\\\bm w_1\neq\bm w_2}}
\frac{d(\widehat g_{2:J+1,T_0+h}^{(\bm w_1)},\widehat g_{2:J+1,T_0+h}^{(\bm w_2)})}{\|\bm w_1-\bm w_2\|_1^\kappa}=O_p(1)
\]
holds for some fixed $\kappa\in(0,1]$. Then
$d(\widehat g_{2:J+1,T_0+h}^{(\widehat{\bm w})},
\widehat g_{2:J+1,T_0+h}^{(\widetilde{\bm w})})=o_p(1)$ and, with $t=T_0+h$,
\begin{align}\label{ineq:GSCM-bound}
&d(Y_{1,t}(N),\widehat Y_{1,t}^{(\mathrm{GSC})}) \nonumber\\
&\le
d(g_t(\mathcal P_{1,t}(U_1)),g_t(U_1))
+d(g_t(U_1),g_{2:J+1,t}^{(\widetilde{\bm w})})
+d(g_{2:J+1,t}^{(\widetilde{\bm w})},
\widehat g_{2:J+1,t}^{(\widetilde{\bm w})})+o_p(1).
\end{align}
\item[(ii)] Suppose, in addition, that Assumption~\ref{ass:GSC1}(ii) holds and
$\widetilde{\mathcal W}\subset\bigcap_{s=1}^{T_0}\mathbb W_s^*$ for all sufficiently large $T_0$.
Then, for every fixed $h\geq1$,
\[
d(g_{T_0+h}(U_1),g_{2:J+1,T_0+h}^{(\widetilde{\bm w})})=0
\]
for all sufficiently large $T_0$.
\end{itemize}
\end{theorem}

Theorem~\ref{thm:GSC-validity} establishes consistency of the estimated weights for $\widetilde{\mathcal W}$ and decomposes the post-treatment prediction error into the treated-outcome perturbation, latent-signal imbalance, synthetic-control perturbation, and an asymptotically negligible weight-estimation term. Part~(ii) eliminates the latent-signal term when every population-target weight also balances the pre-treatment latent signal and this balance persists after treatment. Corollary~S1 of the Supplement gives sufficient conditions for consistent recovery of the untreated counterfactual.

\begin{remark}[{\it Geodesic synthetic control method  with covariates}]
Suppose covariates $X_{j,t}\in\mathcal M_X$, $j=1,\ldots,J+1$ and $t=1,\ldots,T_0$, are available, where $(\mathcal{M}_X,d_X)$ may differ from the outcome space $(\mathcal{M},d)$. Covariates can be incorporated by augmenting the pre-treatment matching criterion. For example, for a scale parameter $\eta\geq0$, one may obtain weights
\[
\widehat{\bm{w}}(\eta)\in\argmin_{\bm{w}\in\Delta^{J-1}}\left\{\frac{1}{T_0}\sum_{t=1}^{T_0}d^2(Y_{1,t},Y_{2:J+1,t}^{(\bm{w})})+\eta\frac{1}{T_0}\sum_{t=1}^{T_0}d_X^2(X_{1,t},X_{2:J+1,t}^{(\bm{w})})\right\},
\]
where
\[
Y_{2:J+1,t}^{(\bm{w})}=\argmin_{\nu\in\mathcal{M}}\sum_{j=2}^{J+1}w_jd^2(\nu,Y_{j,t}),\quad X_{2:J+1,t}^{(\bm{w})}=\argmin_{\nu\in\mathcal{M}_X}\sum_{j=2}^{J+1}w_jd_X^2(\nu,X_{j,t}).
\]
The corresponding covariate-adjusted GSC estimator is
\[
\widehat Y_{1,t}^{(\mathrm{COV})}=\argmin_{\nu\in\mathcal{M}}\sum_{j=2}^{J+1}\widehat{w}_j(\eta)d^2(\nu,Y_{j,t}),\quad \widehat \tau_t^{(\mathrm{COV})}=\gamma_{\widehat Y_{1,t}^{(\mathrm{COV})},Y_{1,t}},\quad t=T_0+1,\ldots,T.
\]
This formulation balances both pre-treatment outcomes and covariates and reduces to the usual Euclidean SCM construction when both spaces are Euclidean. If multiple covariates are available, one may work in the product space $\mathcal{M}_X=\mathcal{M}_1\times\cdots\times\mathcal{M}_K$ with product metric $d_X^2=d_1^2+\cdots+d_K^2$. Unlike in the Euclidean setting, object-valued outcomes and covariates need not admit a common vector representation, so the metric-space formulation balances each component through its own intrinsic geometry.
\end{remark}

\section{Inference by metric conformal calibration}\label{sec:inf}
We next develop an inference procedure that calibrates prediction errors over time for the treated unit. The procedure is inspired by counterfactual prediction and conformal inference ideas for Euclidean synthetic control methods \citep{ChWuZh21b}, but is formulated intrinsically in the metric space and does not require linear structure.

\subsection{Pointwise prediction regions and tests}\label{subsec:pw}
We first consider pointwise inference for the untreated counterfactual outcome $Y_{1,t}(N)$ at a fixed post-treatment time $t>T_0$. The main idea is to use pre-treatment prediction errors of the treated unit to calibrate the uncertainty of the GSC counterfactual prediction. To avoid using in-sample residuals for calibration, we split the pre-treatment periods into a training set and a calibration set.

Let $\mathcal T_{\mathrm{tr}}\subset \{1,\ldots,T_0\}$ and $\mathcal T_{\mathrm{cal}}\subset \{1,\ldots,T_0\}\setminus \mathcal T_{\mathrm{tr}}$ denote the training and calibration periods, respectively. Using only the training periods, define the GSC training weights
\[
\widehat{\bm w}^{\mathrm{tr}}\in\argmin_{\bm w\in\Delta^{J-1}}\frac{1}{|\mathcal T_{\mathrm{tr}}|}\sum_{s\in\mathcal T_{\mathrm{tr}}}d^2(Y_{1,s},Y^{(\bm w)}_{2:J+1,s}),
\]
where, for any $\bm w=(w_2,\ldots,w_{J+1})^\top\in\Delta^{J-1}$,
\[
Y^{(\bm w)}_{2:J+1,s}=\argmin_{\nu\in\mathcal M}\sum_{j=2}^{J+1}w_j d^2(\nu,Y_{j,s}).
\]
To distinguish the training-based prediction from the full-pre-treatment GSC prediction in Section~\ref{sec:met}, write
\[
\widehat Y^{(\mathrm{GSC,tr})}_{1,s}=\argmin_{\nu\in\mathcal M}\sum_{j=2}^{J+1}\widehat w^{\mathrm{tr}}_j d^2(\nu,Y_{j,s}),\qquad s=1,\ldots,T.
\]

For any time $s$, define the latent no-treatment prediction error $R_s=d(Y_{1,s}(N),\widehat Y^{(\mathrm{GSC,tr})}_{1,s})$. This score is intrinsic: it is a real-valued discrepancy measured directly by the metric $d$ on $\mathcal M$, and  does not require subtraction of random objects or a linear  embedding. 
The residual geodesic from $\widehat Y^{(\mathrm{GSC,tr})}_{1,s}$ to $Y_{1,s}(N)$ contains directional information, but residual geodesics with different endpoints have no canonical ordering in a general geodesic metric space and therefore cannot be ranked directly for conformal calibration. We use their lengths for calibration and retain directional uncertainty through the geodesic effect set defined below.
For pre-treatment periods $s\le T_0$, the treated unit is untreated, so $Y_{1,s}(N)=Y_{1,s}$ and the score is observed. For post-treatment periods $s>T_0$, the score involves the unobserved counterfactual $Y_{1,s}(N)$ and is therefore not observed unless a sharp null hypothesis is imposed.

We use the calibration scores $R_s=d(Y_{1,s},\widehat Y^{(\mathrm{GSC,tr})}_{1,s})$, $s\in\mathcal T_{\mathrm{cal}}$, to estimate the typical size of the GSC prediction error under no treatment. Let $m=|\mathcal T_{\mathrm{cal}}|$, $k_\alpha=\left\lceil (m+1)(1-\alpha)\right\rceil$, and $R_{(k)}$ denote the $k$-th order statistic of $\{R_s:s\in\mathcal T_{\mathrm{cal}}\}$. Define $q_{1-\alpha}=R_{(k_\alpha)}$, with the convention that $q_{1-\alpha}=+\infty$ if $k_\alpha>m$. For a post-treatment period $t>T_0$, the $(1-\alpha)$ prediction region for the untreated counterfactual object is
\[
\mathcal C_t^{(\mathrm{GSC})}(1-\alpha)=\left\{\omega\in\mathcal M:d(\omega,\widehat Y^{(\mathrm{GSC,tr})}_{1,t})\le q_{1-\alpha}\right\}.
\]
Thus $\mathcal C_t^{(\mathrm{GSC})}(1-\alpha)$ is the closed metric ball centered at the GSC counterfactual prediction $\widehat Y^{(\mathrm{GSC,tr})}_{1,t}$ with radius calibrated by held-out pre-treatment prediction errors. Large values of $d(Y_{1,t},\widehat Y^{(\mathrm{GSC,tr})}_{1,t})$ relative to this calibration distribution provide evidence that the observed treated outcome is incompatible with the no-treatment counterfactual trajectory.

\begin{assumption}[Exchangeability of calibration and target scores]
\label{ass:score-exchangeability-pointwise}
Fix a post-treatment time point $t \in \{T_0+1,\ldots,T\}$. Conditional on the data used to estimate the GSC weights, the calibration scores $\{R_s:s\in\mathcal T_{\mathrm{cal}}\}$ together with the target score $R_t$ are exchangeable.
\end{assumption}

Assumption~\ref{ass:score-exchangeability-pointwise} concerns temporal invariance of the prediction-error scores under the no-treatment law, conditional on the training data. It does not require independence across units. The following result gives concrete sufficient conditions in terms of the perturbation vectors and the time maps in model~\eqref{eq:GSC-model}.

\begin{proposition}[Sufficient conditions for score exchangeability]\label{prp:score-exchangeability}
Let $\mathcal F_{\mathrm{tr}}$ denote the $\sigma$-field generated by the data used to estimate $\widehat{\bm w}^{\mathrm{tr}}$, and, for each calibration or target period $s$, define $Z_{j,s}=\mathcal P_{j,s}(U_j)$ and
\[
Z_s=(Z_{1,s},\ldots,Z_{J+1,s}).
\]
Suppose that, conditional on $\mathcal F_{\mathrm{tr}}$, the vectors $Z_s$ are exchangeable over the calibration and target periods. Further assume that either $g_s=g$ throughout these periods or $g_s=h_s\circ g$, where each $h_s:\mathcal M\to\mathcal M$ is a deterministic bijective isometry applied to every unit, and that the relevant weighted Fr\'echet mean map is unique and measurable. Then Assumption~\ref{ass:score-exchangeability-pointwise} holds.
\end{proposition}

This proposition permits arbitrary contemporaneous dependence among the components of each $Z_s$. For calibration and target blocks of equal length, conditional exchangeability of the corresponding block vectors formed from $\{Z_s\}$ similarly implies Assumption~\ref{ass:block-exchangeability} below. In particular, conditional independence and identical distribution of $Z_s$ across the relevant periods are sufficient for both assumptions. Exchangeability of the perturbation vectors alone, however, is insufficient if unrestricted temporal variation in $g_s$ changes the distribution of the resulting scores. When exact exchangeability is implausible, one may instead impose stationarity and weak-dependence conditions directly on the metric prediction-error process and use moving, overlapping, or blockwise calibration schemes, following \citet{ChWuZh21b}. Under the corresponding temporal conditions, these procedures provide approximate or asymptotic validity rather than the exact finite-sample guarantee established in Theorem~\ref{thm:pointwise-conformal}.

\begin{theorem}[Finite-sample pointwise coverage]
\label{thm:pointwise-conformal}
Suppose Assumption~\ref{ass:score-exchangeability-pointwise} holds. Then, for any $t\in\{T_0+1,\ldots,T\}$, $\Pr\left\{Y_{1,t}(N)\in \mathcal C_t^{(\mathrm{GSC})}(1-\alpha)\right\}\ge 1-\alpha$. Consequently, under the pointwise null hypothesis of no treatment effect $H_{0,t}:Y_{1,t}(I)=Y_{1,t}(N)$, the test
\[
\phi_t=1\left\{Y_{1,t}\notin \mathcal C_t^{(\mathrm{GSC})}(1-\alpha)\right\}=1\left\{d\left(Y_{1,t},\widehat Y_{1,t}^{(\mathrm{GSC,tr})}\right)>q_{1-\alpha}\right\}
\]
has finite-sample size at most $\alpha$.
\end{theorem}

The pointwise test can equivalently be reported through the conformal rank $p$-value
\[
\widehat p_t=\frac{1+\sum_{s\in\mathcal T_{\mathrm{cal}}}1\{R_s\geq d(Y_{1,t},\widehat Y_{1,t}^{(\mathrm{GSC,tr})})\}}{m+1},
\]
with smaller values providing stronger evidence against $H_{0,t}$. Under $H_{0,t}$, Assumption~\ref{ass:score-exchangeability-pointwise} implies $\Pr(\widehat p_t\leq\alpha)\leq\alpha$; the non-strict comparison is conservative in the presence of ties.

Theorem~\ref{thm:pointwise-conformal} provides a finite-sample valid prediction region for the untreated counterfactual random object $Y_{1,t}(N)$. Since the causal estimand in our framework is the geodesic $\tau_t=\gamma_{Y_{1,t}(N),Y_{1,t}(I)}$, the prediction region for $Y_{1,t}(N)$ naturally induces uncertainty quantification for the geodesic treatment effect itself. Specifically, for $t>T_0$, define
\[
\mathcal G_t(1-\alpha)=\left\{\gamma_{\omega,Y_{1,t}}:\omega\in \mathcal C_t^{(\mathrm{GSC})}(1-\alpha)\right\}.
\]
Because the observed post-treatment outcome satisfies $Y_{1,t}=Y_{1,t}(I)$ for $t>T_0$, the set $\mathcal G_t(1-\alpha)$ may be interpreted as a confidence set for the geodesic causal effect.

\begin{corollary}[Confidence set for the geodesic effect]
\label{cor:geodesic-effect-set}
Suppose the assumptions of Theorem~\ref{thm:pointwise-conformal} hold. Then, for any $t\in\{T_0+1,\ldots,T\}$, $\Pr\left\{\tau_t\in \mathcal G_t(1-\alpha)\right\}\ge 1-\alpha$.
\end{corollary}

In addition to the geodesic effect itself, a natural scalar summary is the magnitude of the treatment effect, $\delta_t=d(Y_{1,t}(N),Y_{1,t}(I))$, which corresponds to the geodesic distance between the untreated and treated potential outcomes. The corresponding estimator is $\widehat\delta_t=d(\widehat Y_{1,t}^{(\mathrm{GSC,tr})},Y_{1,t})$. The conformal prediction region immediately yields a finite-sample confidence interval for $\delta_t$.

\begin{corollary}[Confidence interval for effect size]
\label{cor:effect-magnitude}
Suppose the assumptions of Theorem~\ref{thm:pointwise-conformal} hold. Then, for any $t\in\{T_0+1,\ldots,T\}$, $\Pr\left\{\delta_t\in\left[\{\widehat\delta_t-q_{1-\alpha}\}_+,\,\widehat\delta_t+q_{1-\alpha}
\right]\right\}\ge 1-\alpha$, where $\{a\}_+=\max(a,0)$.
\end{corollary}

Corollary~\ref{cor:effect-magnitude} shows that uncertainty quantification for the untreated counterfactual outcome directly translates into uncertainty quantification for the magnitude of the geodesic treatment effect through the metric structure of the outcome space. Unlike for the special case of Euclidean synthetic control methods, the proposed inference procedure is formulated intrinsically in the metric space. 

\subsection{A global no-effect test}

We now develop a global test for detecting whether the treatment induces any departure from the untreated trajectory over the entire post-treatment period. Specifically, consider the null hypothesis
\[
H_0^{\mathrm{glob}}:Y_{1,t}(I)=Y_{1,t}(N),\quad t=T_0+1,\ldots,T.
\]
Under $H_0^{\mathrm{glob}}$, the observed post-treatment trajectory coincides with the untreated potential trajectory, so deviations between the observed treated unit and its geodesic synthetic control should resemble the treated unit's own pre-treatment prediction errors.

Let $\mathcal B_*=\{T_0+1,\ldots,T\}$ denote the post-treatment block and let $T_1=T-T_0$ denote the number of post-treatment periods. To calibrate the post-treatment prediction error, partition the remaining pre-treatment periods into disjoint calibration blocks $\mathcal B_1,\ldots,\mathcal B_K\subset\{1,\ldots,T_0\}\setminus\mathcal T_{\mathrm{tr}}$, each of length $T_1$. For a block $\mathcal B$ and $r\in[1,\infty)$, define the block score
\[
S_r(\mathcal B)=\left\{\frac{1}{|\mathcal B|}\sum_{t\in\mathcal B}d^r\left(Y_{1,t},\widehat Y_{1,t}^{(\mathrm{GSC,tr})}\right)\right\}^{1/r},
\]
and for $r=\infty$, define $S_\infty(\mathcal B)=\max_{t\in\mathcal B}d\left(Y_{1,t},\widehat Y_{1,t}^{(\mathrm{GSC,tr})}\right)$.

The choice of $r$ determines the sensitivity of the test to different alternatives. The statistic $S_\infty$ is particularly sensitive to large but localized treatment effects, whereas smaller values such as $r=1$ or $r=2$ are more sensitive to persistent deviations from the synthetic counterfactual trajectory.

Under $H_0^{\mathrm{glob}}$, the post-treatment block score is observable and given by
\[
S_r(\mathcal B_*)=\left\{\frac{1}{T_1}\sum_{t=T_0+1}^{T}d^r\left(Y_{1,t},\widehat Y_{1,t}^{(\mathrm{GSC,tr})}\right)\right\}^{1/r},
\]
with the obvious modification when $r=\infty$. We then define the conformal $p$-value
\[
\widehat p_r=\frac{1+\sum_{k=1}^K1\{S_r(\mathcal B_k)\ge S_r(\mathcal B_*)\}}{K+1}.
\]

\begin{assumption}[Exchangeability of block scores under no treatment]
\label{ass:block-exchangeability}
Conditional on the data used to estimate the GSC weights, and under $H_0^{\mathrm{glob}}$, the block scores $S_r(\mathcal B_1),\ldots,S_r(\mathcal B_K)$, $S_r(\mathcal B_*)$ are exchangeable.
\end{assumption}

Assumption~\ref{ass:block-exchangeability} requires that, under the no-treatment law, the treated unit's metric prediction errors remain distributionally invariant at the block level. As in the pointwise setting, this condition concerns temporal stability and does not require exchangeability or independence across units. Proposition~\ref{prp:score-exchangeability} gives sufficient conditions.

\begin{theorem}[Finite-sample validity of the global test]
\label{thm:global-test}
Suppose Assumption~\ref{ass:block-exchangeability} holds. Then the test $\phi_r^{\mathrm{glob}}=1\{\widehat p_r\le\alpha\}$ satisfies $\Pr_{H_0^{\mathrm{glob}}}\left\{\phi_r^{\mathrm{glob}}=1\right\}\le\alpha$.
\end{theorem}

Theorem~\ref{thm:global-test} establishes finite-sample validity of the proposed global test. The procedure may be interpreted as an end-of-sample stability test for the treated unit's intrinsic prediction errors. Rejection of $H_0^{\mathrm{glob}}$ indicates that the post-treatment trajectory lies systematically farther from its geodesic synthetic counterfactual than would be expected based on the treated unit's own pre-treatment prediction errors.

\begin{remark}[In-time placebo diagnostic]\label{rem:intime-placebo}
An important diagnostic for synthetic control analyses is an in-time placebo, or backdating, check \citep{abad:15}. Choose a pseudo-treatment time $\widetilde T_0<T_0$, estimate the GSC weights using only periods $t\leq \widetilde T_0$, and construct pseudo-post-treatment synthetic controls for $\widetilde T_0<t\leq T_0$. Since no intervention occurred during this period, the discrepancies $d\left(Y_{1,t},\widehat Y_{1,t}^{(\mathrm{GSC},\widetilde T_0)}\right)$, $\widetilde T_0<t\leq T_0$, should be small if the counterfactual prediction is stable. Large pseudo-post-treatment discrepancies indicate potential lack of fit, anticipation effects, or instability of the donor pool. This diagnostic is the metric analogue of the in-time placebo check commonly used in Euclidean SCM, with the usual residuals replaced by intrinsic metric discrepancies.
\end{remark}

\section{Implementation and simulations}\label{sec:sim}
\subsection{Implementation details}
Algorithm~\ref{alg:gsc} outlines the implementation steps for the proposed geodesic synthetic control method. The optimization problem in Step 1 of Algorithm~\ref{alg:gsc} requires customized approaches that depend on the specific metric and geometry of the underlying metric space. For networks (Example~\ref{exm:net}), SPD matrices (Example~\ref{exm:cov}), distributions with the Wasserstein metric (Example~\ref{exm:mea}), and functions (Example~\ref{exm:fun}), the optimization reduces to a straightforward convex quadratic optimization problem that can be easily implemented \citep{stel:20}.

For the case of compositional data (Example~\ref{exm:com}), the optimization becomes more challenging, as it involves minimizing a bilevel objective where the weighted \f mean $Y_{2:J+1,t}^{(\bm{w})}$ is itself an argmin problem. Due to the complexity of analytical differentiation when nonlinear \f means are involved,  we adopt a derivative-free constrained optimization strategy. Specifically, we use the Constrained Optimization BY Linear Approximations (COBYLA) algorithm \citep{powe:94}, as implemented in the R package \texttt{nloptr} \citep{john:08:1}. The weighted \f mean involved in the objective function is obtained using the R package \texttt{manifold} \citep{mull:20:5}. Further computational details, including explicit formulations of the optimization problems and solver strategies across different geodesic metric spaces, are provided in Section~S.1 of the Supplement.

\begin{algorithm}
    \single
    \KwIn{data $\{Y_{j, t}: j=1, \dots, J+1, t=1, \ldots, T\}$.}
    \KwOut{geodesic synthetic control estimator $\widehat{\tau}_t^{(\mathrm{GSC})}$ for $t=T_0+1, \ldots, T$.}
    $\widehat{\bm{w}}=(\widehat{w}_2,\dots,\widehat{w}_{J+1})\longleftarrow$ the estimated synthetic control weights:
    \[\widehat{\bm{w}}\in\argmin_{\bm{w} \in \Delta^{J-1}}\frac{1}{T_0}\sum_{t=1}^{T_0}d^2(Y_{1,t},Y_{2:J+1,t}^{(\bm{w})})
    \]
    where $Y_{2:J+1,t}^{(\bm{w})}=\argmin_{\nu \in \mathcal{M}}\sum_{j=2}^{J+1}w_jd^2(\nu, Y_{j,t})$ for $t=1, \ldots, T_0$\;
    $\widehat{Y}_{1,t}^{(\mathrm{GSC})}=\argmin_{\nu \in \mathcal{M}}\sum_{j=2}^{J+1}\widehat{w}_jd^2(\nu, Y_{j,t})\longleftarrow$ the synthetic control unit\;
    $\widehat{\tau}_t^{(\mathrm{GSC})}=\gamma_{\widehat{Y}_{1,t}^{(\mathrm{GSC})}, Y_{1,t}}\longleftarrow$ the geodesic synthetic control estimator.
    \caption{Geodesic Synthetic Control Method}
\label{alg:gsc}
\end{algorithm}

\subsection{Simulations for random-object outcomes}
We evaluate network outcomes represented by graph Laplacians of weighted 10-node two-block networks under the Frobenius metric; parallel SPD results are reported in Section~S.5 of the Supplement. The perturbed geodesic model~\eqref{eq:GSC-model-g} uses $\alpha_t=.5$, $J=20$ donors, within- and between-community edge probabilities $.75$ and $.10$, and nonzero latent edge weights drawn from $\mathrm{Unif}(.5,1.5)$. We draw oracle weights $\bm w^*\in\Delta^{J-1}$, set the treated latent network to the corresponding donor combination, and use a common periodic reference network. Each nonzero edge weight is multiplied independently across edges, units, and time by $\exp(\sigma\xi)$, where $\xi\sim\mathrm{Unif}[-\sqrt{3},\sqrt{3}]$, and we set $\sigma=.10$.

For estimation, we use $T_0\in\{20,100,500\}$, one post-treatment period, and 500 replications. Table~\ref{tab:sim} reports the post-treatment prediction gap $d(\widehat g_{2:J+1,T_0+1}^{(\widehat{\bm w})}, \widehat g_{2:J+1,T_0+1}^{(\widetilde{\bm w})})$, the realized post-treatment error $d(Y_{1,T_0+1}(N), \widehat g_{2:J+1,T_0+1}^{(\widehat{\bm w})})$, the weight-estimation error $\|\widehat{\bm w}-\widetilde{\bm w}\|_1$, and the oracle-weight gap $\|\widetilde{\bm w}-\bm w^*\|_1$, where $\widetilde{\bm w}$ denotes the selected minimizer of $\widetilde Q$ under the current simulation design. The post-treatment prediction gap and weight-estimation error decrease with $T_0$, whereas the realized post-treatment error and oracle-weight gap remain stable, consistent with the decomposition in Theorem~\ref{thm:GSC-validity}. These patterns have direct analogues in Euclidean synthetic control: idiosyncratic disturbances can lead the population criterion to select weights that differ from latent-balancing weights \citep{FePi:21}, while a newly realized post-treatment disturbance remains in the counterfactual prediction error \citep{AbDiHa10}. Increasing $T_0$ therefore improves estimation of the population target but need not eliminate the oracle-weight gap or newly realized post-treatment disturbances.

\begin{table}[tb]
\single
\centering
\begin{tabular}{lccc}
\hline
Quantity & $T_0=20$ & $T_0=100$ & $T_0=500$ \\
\hline
Post-treatment prediction gap & .068 (.001) & .031 ($<$.001) & .014 ($<$.001) \\
Realized post-treatment error & .374 (.003) & .373 (.003) & .367 (.003) \\
$\|\widehat{\bm w}-\widetilde{\bm w}\|_1$ & .120 (.001) & .055 (.001) & .025 ($<$.001) \\
$\|\widetilde{\bm w}-\bm w^*\|_1$ & .063 (.001) & .062 (.001) & .062 (.001) \\
\hline
\end{tabular}
\caption{Network estimation results for $T_0\in\{20,100,500\}$ and $\sigma=.10$, based on 500 replications. The post-treatment prediction gap compares the post-treatment predictions based on $\widehat{\bm w}$ and $\widetilde{\bm w}$, while the realized post-treatment error compares the untreated post-treatment outcome with the prediction based on $\widehat{\bm w}$. The remaining rows report the $\ell_1$ weight-estimation error and oracle-weight gap. Entries are Monte Carlo means with Monte Carlo standard errors in parentheses.}
\label{tab:sim}
\end{table}

Section~S.5 of the Supplement varies $T_0$ and $\sigma$ separately, showing that longer pre-treatment periods reduce the weight-estimation error and post-treatment prediction gap and that larger perturbations increase realized post-treatment error.

\begin{figure}[tb]
\single
\centering
\includegraphics[width=.6\linewidth]{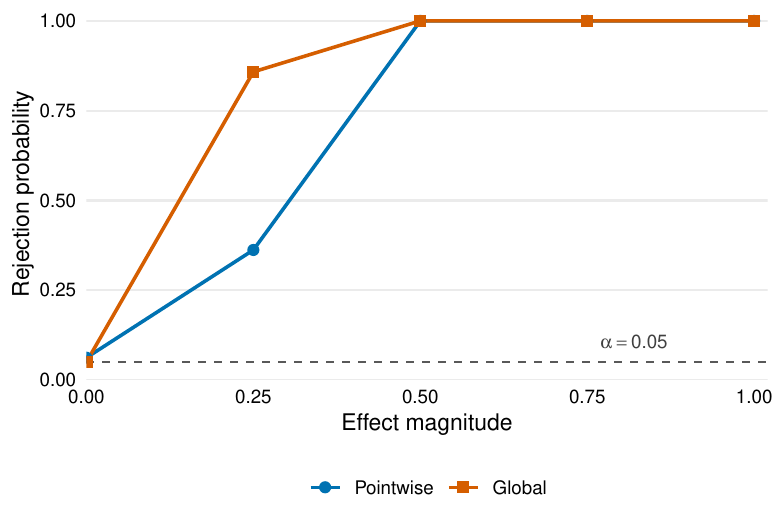}
\caption{Empirical rejection probabilities for the pointwise and global tests in the network design over 500 replications. The dashed line marks the nominal level $\alpha=.05$; values at zero effect report empirical size.}
\label{fig:sim-network-inference}
\end{figure}

For inference, $T_0=500$ is split into 200 training and 300 calibration periods, followed by five post-treatment periods. Treatment adds $a$ times a normalized between-community edge Laplacian, giving effect magnitude $a\in\{0,.25,.50,.75,1\}$. Pointwise inference uses the first post-treatment period; the global test uses 60 five-period calibration blocks with the block score set to $r=2$. The common time component cancels from each prediction error because the weights sum to one, so the i.i.d. perturbations make the calibration and target scores exchangeable conditional on training. Under the null, pointwise coverage is $.938$ and pointwise and global rejection probabilities are $.062$ and $.050$. At $a=.25$, the corresponding powers are $.362$ and $.858$, and both reach one at $a=.50$; see Figure~\ref{fig:sim-network-inference}.

\section{Real data examples}\label{sec:app}

\subsection{Employment composition after the 2011 Great East Japan Earthquake}
We analyze the impact of the 2011 Great East Japan Earthquake on the employment-population ratio by employment category, using data from the Research Institute of Economy, Trade and Industry (RIETI) database (\url{https://www.rieti.go.jp/jp/database/R-JIP2021/index.html}). These data provide annual employment figures by industry for each prefecture in Japan from 1994 to 2018. Employment is categorized into three broad groups: primary (agriculture, forestry, and fisheries), secondary (manufacturing, electricity, gas, water supply, waste management and construction) and tertiary (services, including wholesale/retail, finance, and healthcare). For each prefecture and each year, the employment shares (proportions)  across these three categories are represented as a three-dimensional compositional vector. The component-wise square root transformation (see Example~\ref{exm:com}) maps these compositional outcomes onto the positive orthant of the sphere $\mathcal{S}_+^2 \subset \mathbb{R}^3$, which we endow with the geodesic metric. Unlike standard synthetic control approaches that typically analyze each compositional component separately, the proposed approach considers the entire compositional vector jointly. This holistic perspective reflects the intrinsic geometry of compositional data, thereby capturing interdependencies of employment across the three categories.

The Great East Japan Earthquake occurred on March 11, 2011, causing widespread devastation across northeastern Japan, with Miyagi Prefecture experiencing particularly severe damage. The disaster substantially disrupted regional economic activity and induced major structural changes in employment. To evaluate the effect of the earthquake on employment composition, we treat Miyagi Prefecture as the treated unit. Since the earthquake constituted a broad nationwide shock that may also have affected nearby prefectures, we restrict the donor pool to prefectures located in Western Japan, which were geographically distant from the disaster and therefore less likely to have been directly exposed to its economic and social impacts. Specifically, the donor pool consists of the following 20 prefectures: Shiga, Kyoto, Nara, Wakayama, Tottori, Shimane, Okayama, Hiroshima, Yamaguchi, Tokushima, Kagawa, Ehime, Kochi, Fukuoka, Saga, Nagasaki, Kumamoto, Oita, Miyazaki, and Kagoshima. We analyze annual employment composition data from 1994 to 2013, using 1994--2010 as the pre-treatment period and 2011--2013 as the post-treatment period.

\begin{figure}[tb]
    \single
    \centering
    \includegraphics[width=0.75\linewidth]{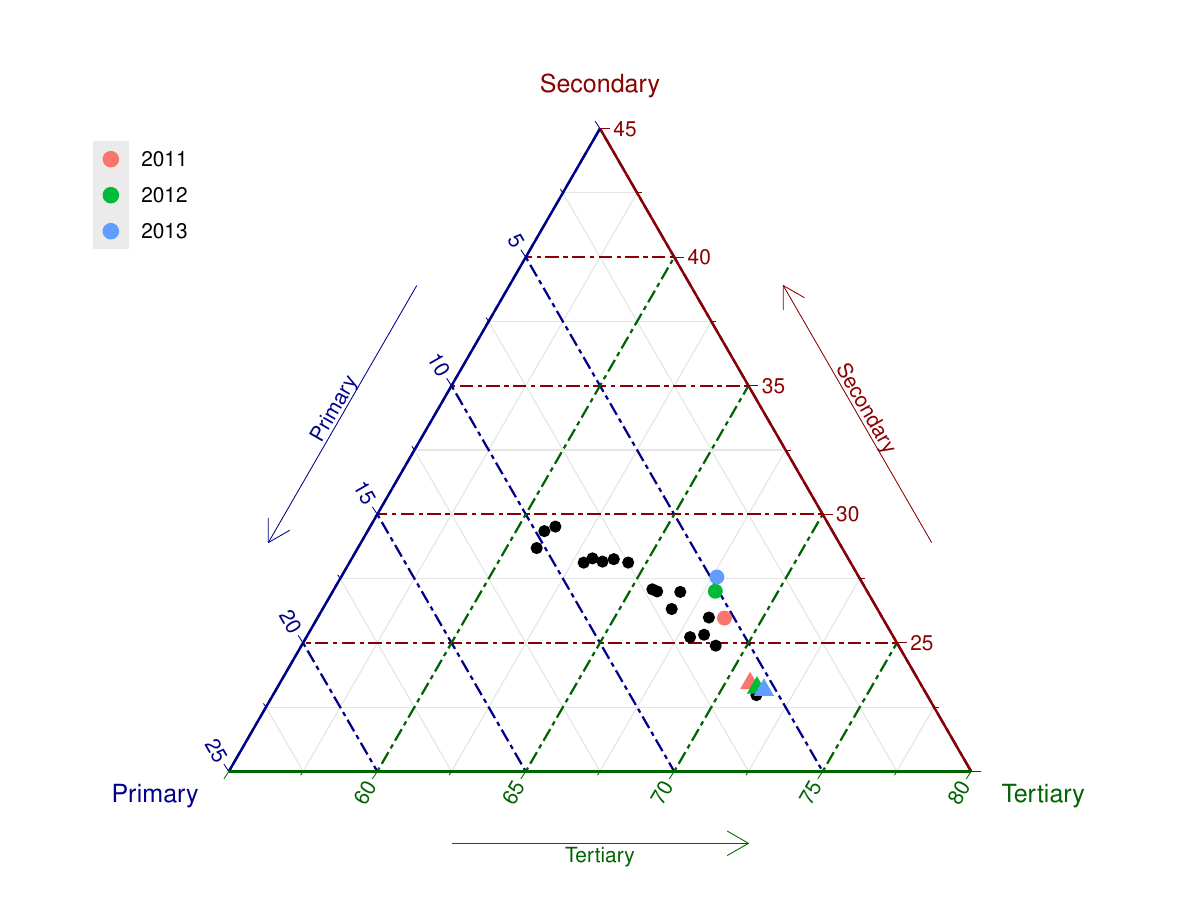}
    \caption{Ternary plot of employment composition in Miyagi Prefecture before and after the 2011 Great East Japan Earthquake. Black points represent the observed pre-treatment trajectory of Miyagi from 1994--2010. Colored circles denote the observed post-treatment compositions for 2011--2013, while colored triangles denote the corresponding geodesic synthetic counterfactuals, with colors indicating year. Relative to the synthetic counterfactual trajectory, the observed post-treatment compositions exhibit a higher proportion in the secondary employment category and a lower proportion in the tertiary category, while the primary category remains comparatively stable.}
    \label{fig:earthquaket}
\end{figure}

We apply the proposed GSC method to estimate the causal effect of the earthquake on employment composition in Miyagi Prefecture. Figure~\ref{fig:earthquaket} displays the observed and synthetic post-treatment employment compositions using a ternary plot. Relative to the synthetic counterfactual trajectory, the observed post-treatment compositions exhibit a larger proportion of secondary category employment and a smaller proportion of tertiary category employment, while the primary category remains comparatively stable.

For comparison, we fit conventional SCM separately to the three employment shares. The distinct donor weights do not preserve the unit-sum constraint: the maximum absolute deviation is $.005249$ over the held-out pre-treatment period and $.002813$ after treatment. GSC enforces this constraint through a single joint objective; Section~S.5 of the Supplement reports the componentwise predictions.

This shift is consistent with reconstruction-driven demand for secondary-category employment and prolonged disruption of service-sector activity. It suggests that the earthquake temporarily changed Miyagi's employment composition toward reconstruction work.

Using 1994--1999 for training and 2000--2010 for calibration, the estimated effect magnitudes for 2011--2013 are $.0421$, $.0561$, and $.0646$, with $90\%$ split-conformal intervals $[.0066,.0776]$, $[.0206,.0916]$, and $[.0291,.1001]$. Each conformal $p$-value is $1/12$; 11 calibration periods do not yield a nontrivial $95\%$ split-conformal interval.

\subsection{Impact of the 1972 abortion legislation on fertility patterns in East Germany}
The study of fertility rates has attracted substantial attention, with increasing emphasis on analyzing age-specific fertility rates (ASFR), encapsulating more information than summary measures such as the total fertility rate \citep{mull:17:4}. Represented as fertility curves across reproductive ages, ASFR provides rich insights into fertility, allowing researchers to detect age-specific causal effects that aggregate statistics cannot capture. 

In March 1972, East Germany introduced legislation permitting women to terminate pregnancies within the first twelve weeks, significantly liberalizing abortion access. Although the primary intention was reproductive autonomy, an unintended consequence was a substantial drop in fertility, with the total fertility rate decreasing notably from 2.1 children per woman in 1971 to 1.5 by 1975 \citep{butt:90}. To gain deeper insights into how this policy affected fertility behavior across different reproductive ages, we applied the proposed  GSC to ASFR curves. When viewed as functions over reproductive ages, these are functional data in the space $L^2$ (see Example~\ref{exm:fun}).

\begin{figure}[!tb]
    \single
    \centering
    \includegraphics[width=.7\linewidth]{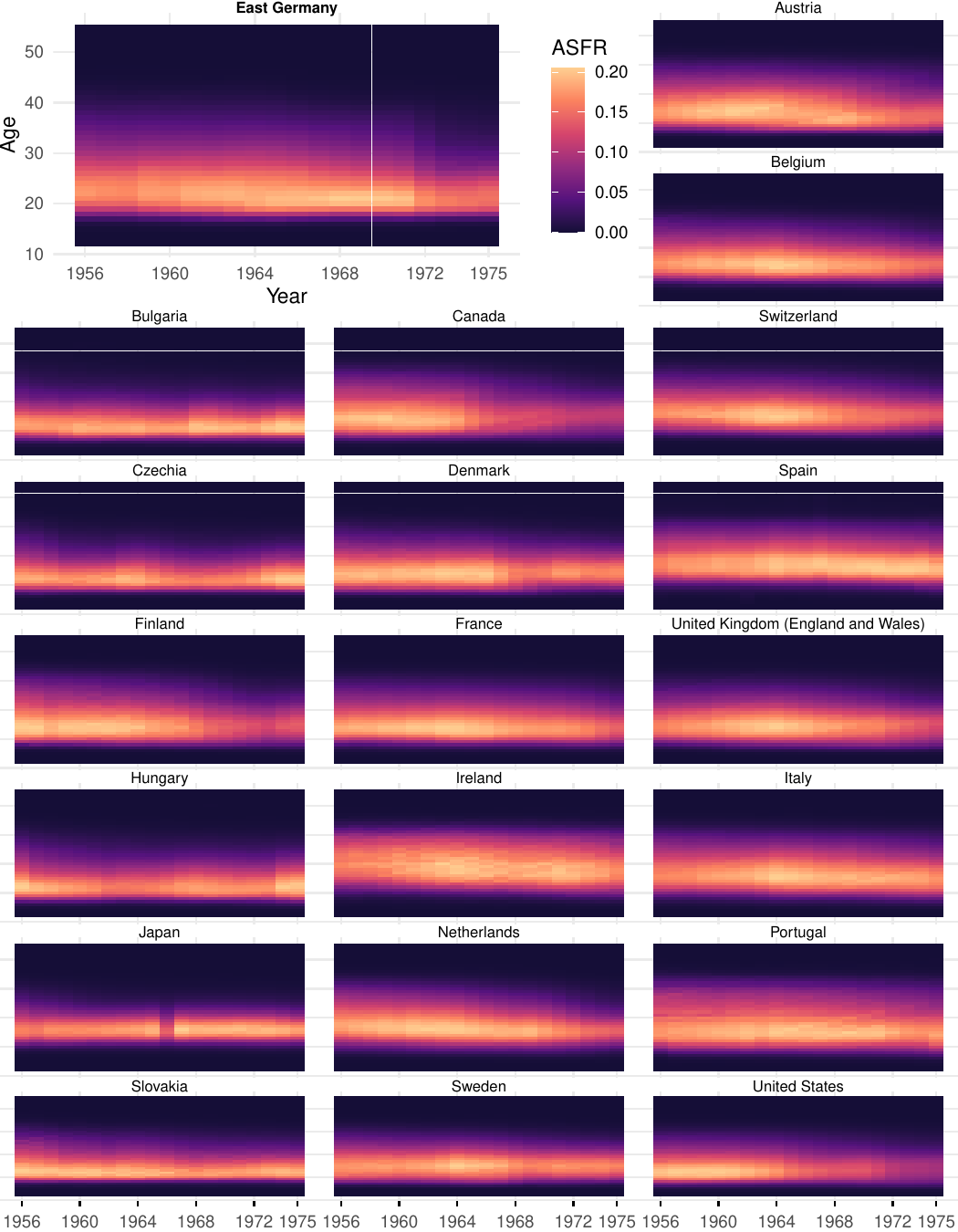}
    \caption{Age-specific fertility rates (ASFR) for East Germany and the control countries, 1956--1975. Each column represents the ASFR across ages 12--55 in the year shown on the horizontal axis.}
    \label{fig:asfr}
\end{figure}

The data for this analysis were obtained from the \citeauthor{HFD}, which provides ASFR annually for ages 12 to 55 for each calendar year across different countries. We select East Germany as the treated unit, given its distinct policy change, and utilize a control group composed of 20 countries that did not experience comparable abortion liberalization during this period (Austria, Belgium, Bulgaria, Canada, Switzerland, Czechia, Denmark, Spain, Finland, France, Hungary, Ireland, Italy, Japan, Netherlands, Portugal, Slovakia, Sweden, United Kingdom (England and Wales), and United States). We define 1956--1971 as pre-treatment periods and 1972--1975 as post-treatment periods, so that it is possible to capture both immediate and medium-term policy impacts. Figure~\ref{fig:asfr} displays the ASFR curves for East Germany and the control countries from 1956 to 1975. A clear reduction is evident for East Germany during the post-treatment period.

\begin{figure}[tb]
    \single
    \centering
    \includegraphics[width=0.9\linewidth]{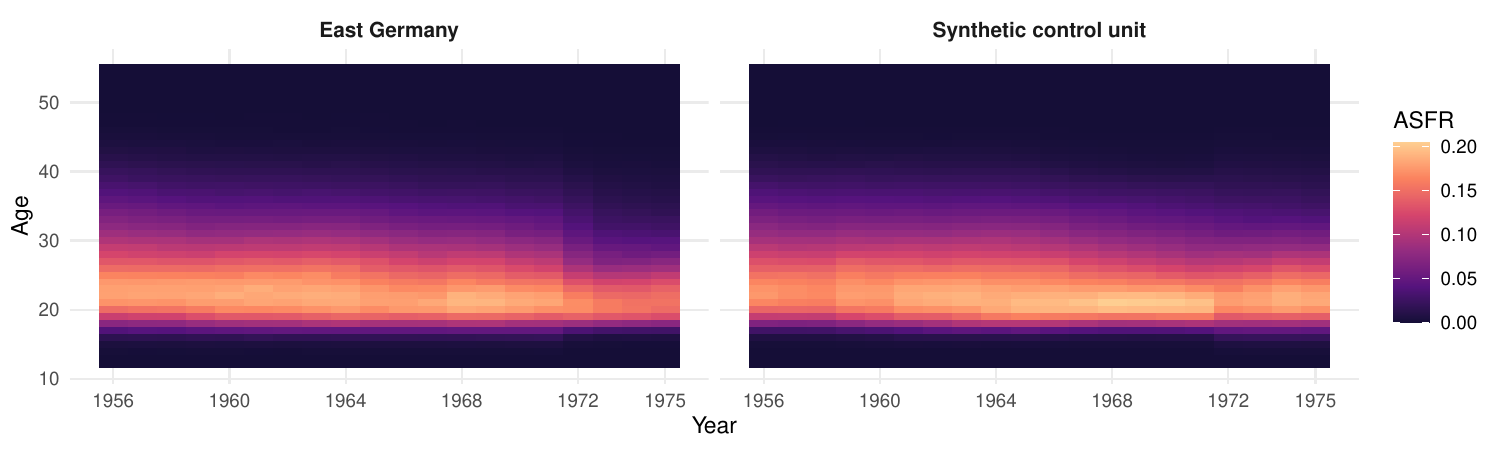}
    \caption{Heat maps of age-specific fertility rates (ASFR) for East Germany (left) and the synthetic control unit (right) from 1956 to 1975. Each column represents the ASFR across ages 12 to 55 (vertical axis) for each calendar year (horizontal axis).}
    \label{fig:asfrsc}
\end{figure}

Figure~\ref{fig:asfrsc} compares the observed ASFR of East Germany with the synthetic control constructed from the donor countries. The synthetic control closely tracks East Germany's pre-treatment fertility patterns, while after 1972 the observed trajectory exhibits a clear downward shift relative to the synthetic counterfactual, especially across prime childbearing ages. Using 1956--1961 for training and 1962--1971 for calibration, the estimated effect magnitudes for 1972--1975 are $.0813$, $.1279$, $.1544$, and $.1332$, with $90\%$ split-conformal intervals $[.0207,.1419]$, $[.0673,.1885]$, $[.0938,.2150]$, and $[.0726,.1938]$. Each conformal $p$-value is $1/11$; 10 calibration periods do not yield a nontrivial $95\%$ split-conformal interval.

As a diagnostic check, we also conduct an in-time placebo analysis by backdating the treatment year to $\widetilde T_0=1967$, as discussed in Remark~\ref{rem:intime-placebo}. Using only pre-1968 observations to estimate the GSC weights, the pseudo-post-treatment discrepancies for 1968--1971 are $0.032$, $0.036$, $0.039$, and $0.042$, respectively. These are substantially smaller than the true post-treatment discrepancies, $0.065$, $0.116$, $0.141$, and $0.122$ for 1972--1975. This diagnostic supports the interpretation that the post-1972 divergence reflects the policy effect rather than a comparable pre-existing discrepancy.

\section{Discussion}\label{sec:dis}
The proposed geodesic synthetic control method extends causal inference to random-object outcomes in geodesic metric spaces by replacing Euclidean weighted averages with weighted Fr\'echet means and representing treatment effects as geodesics connecting untreated and treated potential outcomes. This provides an intrinsic extension of the classical synthetic control framework to a broad range of object-valued outcomes, including distributions, compositions, networks, SPD matrices, and functional data. Under the perturbed causal model, we established consistency of the estimated weights for the perturbed population target, derived a post-treatment prediction-error bound, and gave sufficient conditions for consistent recovery of the untreated counterfactual. We also proposed a metric conformal calibration procedure for uncertainty quantification.

Several extensions are natural. First, when the untreated potential outcome of the treated unit is not well approximated by a weighted Fr\'echet mean of the controls, GSC may suffer from imperfect pre-treatment fit. A possible remedy is an augmented GSC estimator, paralleling the augmented synthetic control method of \citet{BeFeRo21}. Developing such a method requires regression-based bias correction for metric space-valued outcomes, potentially building on Fr\'echet regression \citep{mull:19:6} and geodesic optimal transport regression \citep{zhu:23}. Second, it would be useful to develop a geodesic analogue of synthetic difference-in-differences \citep{ArAtHiImWa21}, combining unit reweighting with time reweighting. Recent work on geodesic difference-in-differences \citep{zhou:25} provides a foundation for this direction.

Metric choice is another important modeling issue for GSC. Many object spaces admit multiple meaningful metrics, and the chosen metric determines the distances, weighted Fr\'echet means, geodesics, synthetic counterfactuals, and treatment-effect magnitudes. Consequently, the metric should be selected based on the scientific question, the geometry of the data type, interpretability, and computational feasibility. For example, SPD matrices may be equipped with the Frobenius, Log-Euclidean, power, or Log-Cholesky metric. When several metrics are plausible, data-adaptive criteria or sensitivity analyses may be useful; see \citet{mull:16:2} for adaptive metric selection for SPD matrices and \citet{mull:24} for a broader discussion. A systematic study of metric selection for GSC, especially in causal inference settings, is an important direction for future work.

\section*{Data Availability Statement}
The employment data are openly available from the \href{https://www.rieti.go.jp/jp/database/R-JIP2021/index.html}{RIETI R-JIP database}, and the fertility data are available from the \href{https://www.humanfertility.org}{Human Fertility Database}.

\section*{Acknowledgments}
The authors thank the editor, associate editor, and three anonymous referees for their valuable and constructive feedback.

\section*{Disclosure Statement}
The authors report there are no competing interests to declare.

\single
\bibliographystyle{rss}
\bibliography{collection.bib}
\double

\clearpage
\begin{center}
{\large\bf SUPPLEMENTARY MATERIAL}
\end{center}

\setcounter{section}{0}
\setcounter{subsection}{0}
\setcounter{equation}{0}
\setcounter{figure}{0}
\setcounter{table}{0}
\setcounter{algocf}{0}
\setcounter{proposition}{0}
\setcounter{theorem}{0}
\setcounter{corollary}{0}
\setcounter{assumption}{0}
\setcounter{definition}{0}
\setcounter{remark}{0}

\renewcommand{\thesection}{S.\arabic{section}}
\renewcommand{\thesubsection}{S.\arabic{section}.\arabic{subsection}}
\renewcommand{\theequation}{S\arabic{equation}}
\renewcommand{\thefigure}{S\arabic{figure}}
\renewcommand{\thetable}{S\arabic{table}}
\renewcommand{\thealgocf}{S\arabic{algocf}}

\counterwithout*{assumption}{section}
\counterwithout*{proposition}{section}
\counterwithout*{theorem}{section}
\counterwithout*{corollary}{section}
\counterwithout*{definition}{section}
\counterwithout*{remark}{section}
\renewcommand{\theproposition}{S\arabic{proposition}}
\renewcommand{\thetheorem}{S\arabic{theorem}}
\renewcommand{\thecorollary}{S\arabic{corollary}}
\renewcommand{\theassumption}{S\arabic{assumption}}
\renewcommand{\thedefinition}{S\arabic{definition}}
\renewcommand{\theremark}{S\arabic{remark}}

\providecommand{\theHalgocf}{}
\providecommand{\theHproposition}{}
\providecommand{\theHtheorem}{}
\providecommand{\theHcorollary}{}
\providecommand{\theHassumption}{}
\providecommand{\theHdefinition}{}
\providecommand{\theHremark}{}
\renewcommand{\theHsection}{supp.\arabic{section}}
\renewcommand{\theHsubsection}{supp.\arabic{section}.\arabic{subsection}}
\renewcommand{\theHequation}{supp.\arabic{equation}}
\renewcommand{\theHfigure}{supp.\arabic{figure}}
\renewcommand{\theHtable}{supp.\arabic{table}}
\renewcommand{\theHalgocf}{supp.\arabic{algocf}}
\renewcommand{\theHproposition}{supp.\arabic{proposition}}
\renewcommand{\theHtheorem}{supp.\arabic{theorem}}
\renewcommand{\theHcorollary}{supp.\arabic{corollary}}
\renewcommand{\theHassumption}{supp.\arabic{assumption}}
\renewcommand{\theHdefinition}{supp.\arabic{definition}}
\renewcommand{\theHremark}{supp.\arabic{remark}}

\section{Optimization Details}
This section provides additional computational details for the implementation of the geodesic synthetic control (GSC) 
estimator. We present explicit formulations of the optimization problems, describe how weighted Fr\'echet means are computed in different geodesic metric spaces, and discuss the solver strategies adopted in each case.

\subsection{Unit weights}
The GSC relies on optimizing weights attached to observational units rather than features of the objects themselves. Specifically, the \emph{unit weights} $\widehat{\bm{w}} \in \Delta^{J-1}$ form convex combinations of control units. Random objects are then averaged via weighted Fr\'echet means in the respective metric space. Thus, the weights govern the influence of units but do not vary across internal coordinates of the objects (such as quantiles of a distribution). This preserves interpretability in the geodesic extension.

\subsection{General optimization problems}\label{subsec:supp:opt}

The estimation of unit weights in GSC is formulated as a minimization problem involving distances between the treated unit and weighted Fr\'echet means of control units. Specifically, the unit weights $\widehat{\bm{w}} \in \Delta^{J-1}$ are chosen to minimize the average squared distance between the treated outcome $Y_{1,t}$ and the synthetic control outcome $Y^{(\bm{w})}_{2:J+1,t}$ over the pre-treatment period,
\[
\widehat{\bm{w}} \in \argmin_{\bm{w} \in \Delta^{J-1}} \frac{1}{T_0}\sum_{t=1}^{T_0} d^2\!\left(Y_{1,t},\, Y^{(\bm{w})}_{2:J+1,t}\right),
\]
where the synthetic control at time $t$ is defined through the weighted Fr\'echet mean
\[
Y^{(\bm{w})}_{2:J+1,t} = \argmin_{\nu \in \mathcal M} \sum_{j=2}^{J+1} w_j d^2(\nu, Y_{j,t}).
\]

The optimization is carried out over the probability simplex, with the geometry of the underlying metric space determining how the Fr\'echet means are evaluated.

\subsection{Spaces with closed-form weighted Fr\'echet means}

For networks (Example 1), symmetric positive-definite (SPD) matrices (Example 3), distributions (Example 4), and functional data (Example 5), the weighted Fr\'echet mean admits an explicit expression under the metrics considered in this paper. In these cases, the outer optimization problems in Section~\ref{subsec:supp:opt} reduce to convex quadratic programs in $\bm{w}$, 
which can be solved efficiently using standard quadratic programming solvers such as OSQP \citep{stel:20}. This yields numerically stable and fast implementations.

\subsection{Compositional data with the arc-length distance}

For compositional data with the arc-length distance (Example 2), the situation is more complex. When mapped to the positive orthant of the sphere, the weighted Fr\'echet mean has no closed form and must be obtained via iterative algorithms on the manifold. Each evaluation of the objective therefore requires solving an inner Fr\'echet mean problem, which creates a bilevel optimization problem. Because analytical differentiation is intractable in this nonlinear setting, we employ a derivative-free constrained optimization strategy, specifically the COBYLA algorithm \citep{powe:94} as implemented in the \texttt{nloptr} package \citep{john:08:1}. The inner Fr\'echet mean computations are carried out using the \texttt{manifold} package \citep{mull:20:5}.

\subsection{Practical remarks}

For the tractable cases with closed-form weighted Fr\'echet means, optimization is straightforward and stable. For compositional data with the arc-length distance, convergence of the iterative mean calculation can be slow when many observations lie close to zero components, in which case regularization (e.g., adding a small constant to proportions before normalization) improves stability.

\section{Verification of Assumption 3.1}\label{app:model}

This section gives sufficient conditions for Assumption~3.1 under the signal models and metrics treated below: the Frobenius metric for networks, the arc-length metric for compositional data, the Frobenius and Log-Euclidean metrics, the power metric family, and the Log-Cholesky metric for SPD matrices, the 2-Wasserstein metric for univariate distributions, and the $L^2$ metric for functional data. 
For $\bm{w}=(w_2,\dots,w_{J+1})^\top \in \Delta^{J-1}$, define
\[
U_{2:J+1}^{(\bm{w})} = \argmin_{\nu \in \mathcal{M}}\sum_{j=2}^{J+1}w_jd^2(\nu, U_j),\quad g_t^{(\bm{w})}(U_{2:J+1}) = g_{2:J+1,t}^{(\bm{w})}=\argmin_{\nu\in\mathcal{M}} \sum_{j=2}^{J+1} w_j d^2(\nu, g_t(U_j)),
\]
and $\mathbb{W}^*=\{\bm{w} \in \Delta^{J-1}:d(U_1,U_{2:J+1}^{(\bm{w})})=0\}$. Throughout Sections \ref{exm:cor-app}--\ref{exm:fun-app} below, we assume that $\mathbb{W}^*$ is not empty.

\subsection{Networks}\label{exm:cor-app}

Recall that $g_t(U_j) = \gamma_{\mu_t, U_j}(\alpha_t)= \mu_t + \alpha_t(U_j - \mu_t)$, $\alpha_t \in (0,1)$. For $\bm{w} \in \Delta^{J-1}$,
\begin{align}
U_{2:J+1}^{(\bm{w})} &= \sum_{j=2}^{J+1}w_jU_j, \nonumber \\
g_t^{(\bm{w})}(U_{2:J+1}) &
= \sum_{j=2}^{J+1}w_j\left\{\mu_t + \alpha_t(U_j - \mu_t)\right\} = \mu_t + \alpha_t\left\{\sum_{j=2}^{J+1}w_jU_j - \mu_t\right\} \nonumber \\
&= \mu_t + \alpha_t(U_{2:J+1}^{(\bm{w})} - \mu_t) = g_t(U_{2:J+1}^{(\bm{w})}). \label{eq:SPD-ave}
\end{align}
Furthermore,
\begin{align}
d_F^2(g_t(U_1),g_t(U_{2:J+1}^{(\bm{w})}))
&= \mathrm{tr}\left\{(g_t(U_1) - g_t(U_{2:J+1}^{(\bm{w})}))^\top(g_t(U_1) - g_t(U_{2:J+1}^{(\bm{w})}))\right\} \nonumber \\
&= \alpha_t^2\mathrm{tr}\left\{(U_1 - U_{2:J+1}^{(\bm{w})})^\top(U_1 - U_{2:J+1}^{(\bm{w})})\right\} = \alpha_t^2d_F^2(U_1, U_{2:J+1}^{(\bm{w})}), \label{eq:SPD-dist}
\end{align}
where $\mathrm{tr}(A)$ is the trace of the matrix $A$. This yields $\bigcap_{t=1}^{T_0}\mathbb{W}_t^* = \bigcap_{t=T_0+1}^T\mathbb{W}_t^* = \mathbb{W}^*$ and hence Assumption 3.1 holds.

\subsection{Compositional data }\label{exm:com-app}

For compositional data, we verify Assumption~3.1 under the following spherical signal model. Let
\[
g_t(U_j) = \mathrm{Exp}_{\mu_t}(A_t\mathrm{Log}_{U_1}(U_j)), \quad \mu_t \in \mathcal{S}_+^{d-1},
\]
where $\mathrm{Exp}_{\mu}$ and $\mathrm{Log}_{\mu}$ denote the Riemannian exponential and logarithmic maps on the unit sphere. Let $T_{\mu}\mathcal{S}^{d-1} = \{v\in \mathbb{R}^d: \mu^\top v=0\}$ denote the tangent space of the unit sphere $\mathcal{S}^{d-1}$ at $\mu \in \mathcal{S}_+^{d-1}$. Choose a rotation matrix $R_t \in \mathrm{SO}(d)$ such that $R_t U_1 = \mu_t$, and define $A_t = \kappa_t R_t$, where $\kappa_t>0$ is a constant. Assume that $\kappa_t$ is chosen so that $\|A_t\mathrm{Log}_{U_1}(U_j)\|<\pi$ and $\mathrm{Exp}_{\mu_t}(A_t\mathrm{Log}_{U_1}(U_j))\in\mathcal S_+^{d-1}$ for every $j$. These conditions place the tangent vectors in the injectivity domain of $\mathrm{Exp}_{\mu_t}$ and keep the signal in the outcome space. Then $A_t$ maps $T_{U_1}\mathcal{S}^{d-1}$ into $T_{\mu_t}\mathcal{S}^{d-1}$. Indeed, for any $v \in T_{U_1}\mathcal{S}^{d-1}$, we have $U_1^\top v=0$, and therefore $\mu_t^\top (A_t  v)= \kappa_t U_1^\top R_t^\top R_t v = \kappa_t U_1^\top v = 0$. Thus $A_t v \in T_{\mu_t}\mathcal{S}^{d-1}$, and the expression defining $g_t(U_j)$ is well defined.
Note that $g_t(U_1)= \mathrm{Exp}_{\mu_t}(0) = \mu_t$. 
For every $\bm w\in\Delta^{J-1}$ and time point $t$, assume that, almost surely, the discrete distributions assigning masses $w_j$ to the raw latent objects $U_j$, the latent-signal objects $g_t(U_j)$, and the perturbed objects $g_t(\mathcal P_{j,t}(U_j))$ satisfy the conditions of Theorem~1 in \cite{BuFi01}. Then all corresponding weighted \f means are unique, including those required by Assumption~3.1(i). Additional uniqueness conditions for \f means on Riemannian manifolds are given in Remark~2.8 of \cite{ElHu19}. The same necessity-and-sufficiency argument used below, applied at $U_1$, gives
\[
\mathbb W^*=\left\{\bm w\in\Delta^{J-1}:\sum_{j=2}^{J+1}w_j\mathrm{Log}_{U_1}(U_j)=0\right\}.
\]
For $\overline{\mathbb{W}}_t^*=\{\bm{w} \in \Delta^{J-1}: \sum_{j=2}^{J+1}w_j\mathrm{Log}_{g_t(U_1)}(g_t(U_j))=0\}$, one can verify Assumption 3.1(ii) if
\begin{align}\label{eq:GSC-w-set}
\mathbb{W}_t^* = \overline{\mathbb{W}}_t^*.
\end{align}
Indeed, we have
\begin{align*}
\mathbb{W}_t^*&=\overline{\mathbb{W}}_t^*=\{\bm{w} \in \Delta^{J-1}: \sum_{j=2}^{J+1}w_j\mathrm{Log}_{\mu_t}(\mathrm{Exp}_{\mu_t}(A_t\mathrm{Log}_{U_1}(U_j)))=0\}\\
&=\{\bm{w} \in \Delta^{J-1}: A_t\sum_{j=2}^{J+1}w_j\mathrm{Log}_{U_1}(U_j)=0\}=\{\bm{w} \in \Delta^{J-1}: \sum_{j=2}^{J+1}w_j\mathrm{Log}_{U_1}(U_j)=0\} = \mathbb{W}^*.
\end{align*}
This yields $\bigcap_{t=1}^{T_0}\mathbb{W}_t^* = \bigcap_{t=T_0+1}^T\mathbb{W}_t^* = \mathbb{W}^*$ and hence Assumption 3.1(ii) holds.

Next we show \eqref{eq:GSC-w-set}. Specifically, we establish (i) $\mathbb{W}_t^* \subset \overline{\mathbb{W}}_t^*$ and (ii) $\mathbb{W}_t^* \supset \overline{\mathbb{W}}_t^*$.

\noindent
{\bf Proof of (i)}: Fix an arbitrary $\bm{w}=(w_2,\dots,w_{J+1})^\top \in \mathbb{W}_t^*$. By definition, $g_t^{(\bm{w})}(U_{2:J+1})$ is a \f mean of a random element $X$ such that $\Pr\{X=g_t(U_j)\}=w_j$, $j=2,\dots,J+1$. Further, the condition $d(g_t(U_1),g_t^{(\bm{w})}(U_{2:J+1})) = 0$ yields $g_t(U_1)=g_t^{(\bm{w})}(U_{2:J+1})$. Applying Corollary 3 in \cite{le:14}, we then have $\sum_{j=2}^{J+1}w_j\mathrm{Log}_{g_t(U_1)}(g_t(U_j))=0$. This implies $\bm{w} \in \overline{\mathbb{W}}_t^*$ and hence $\mathbb{W}_t^* \subset \overline{\mathbb{W}}_t^*$ holds.

\vspace{5pt}
\noindent
{\bf Proof of (ii)}:  Fix an arbitrary $\bm{w}=(w_2,\dots,w_{J+1})^\top \in \overline{\mathbb{W}}_t^*$. From the definition of $\overline{\mathbb{W}}_t^*$ and the uniqueness of the \f mean, Theorem 1 in \cite{BuFi01} yields $g_t(U_1)=g_t^{(\bm{w})}(U_{2:J+1})$. This implies $\bm{w} \in \mathbb{W}_t^*$ and hence $\mathbb{W}_t^* \supset \overline{\mathbb{W}}_t^*$.

\subsection{SPD matrices}\label{exm:SPD-app}

First, we consider the space of SPD matrices with the Frobenius metric. Standard arguments lead to  \eqref{eq:SPD-ave} and \eqref{eq:SPD-dist}. Hence Assumption 3.1 holds.

Second, we consider the space of SPD matrices with the power metric family defined as $d_{F,p}(A,B) = \|F_p(A) - F_p(B)\|_F$, where $p>0$ is a constant, $\|\cdot\|_F$ is the Frobenius metric, and $F_p:\mathrm{Sym}^+_m\to\mathrm{Sym}^+_m$ is the matrix power map defined by $F_p(A)=A^p=U\Lambda^pU^\top$, where $U\Lambda U^\top$ is the usual spectral decomposition of $A \in \mathrm{Sym}^+_m$. The geodesic with respect to $d_{F,p}$ between $A$ and $B$ is given as $\gamma_{A,B}(t) = [A^p + t(B^p - A^p)]^{1/p}=F_{1/p}(A^p + t(B^p - A^p))$.
Recall that $g_t(U_j) = \gamma_{\mu_t, U_j}(\alpha_t)$. For $\bm{w} \in \Delta^{J-1}$,
\begin{align*}
(U_{2:J+1}^{(\bm{w})})^p &= \sum_{j=2}^{J+1}w_jU_j^p,\\
(g_t^{(\bm{w})}(U_{2:J+1}))^p &= \mu_t^p + \alpha_t\left\{\sum_{j=2}^{J+1}w_j U_j^p - \mu_t^p\right\} = (g_t(U_{2:J+1}^{(\bm{w})}))^p.
\end{align*}
Furthermore,
\begin{align*}
d_{F,p}^2(g_t(U_1),g_t(U_{2:J+1}^{(\bm{w})}))
&= \|(g_t(U_1))^p - (g_t(U_{2:J+1}^{(\bm{w})}))^p\|^2_F\\
&= \alpha_t^2\| U_1^p - \sum_{j=2}^{J+1}w_j U_j^p\|^2_F
= \alpha_t^2d_{F,p}^2(U_1,U_{2:J+1}^{(\bm{w})}).
\end{align*}
This yields $\bigcap_{t=1}^{T_0}\mathbb{W}_t^* = \bigcap_{t=T_0+1}^T\mathbb{W}_t^* = \mathbb{W}^*$ and hence Assumption 3.1 holds.

Third, we consider the space of SPD matrices with the Log-Euclidean metric, $d_{\mathrm{LE}}(A,B) = \|\log (A) - \log (B)\|_F$. The geodesic with respect to $d_{\mathrm{LE}}$ between $A$ and $B$ is $\gamma_{A,B}(t) = \exp(\log A + t(\log B - \log A))$. Recall $g_t(U_j) = \gamma_{\mu_t, U_j}(\alpha_t)$. For $\bm{w} \in \Delta^{J-1}$,
\begin{align*}
\log (U_{2:J+1}^{(\bm{w})}) &= \sum_{j=2}^{J+1}w_j\log (U_j),\\
\log (g_t^{(\bm{w})}(U_{2:J+1})) &= \log (\mu_t) + \alpha_t\left(\sum_{j=2}^{J+1}w_j\log (U_j) - \log (\mu_t)\right) = \log (g_t(U_{2:J+1}^{(\bm{w})})).
\end{align*}
Further,
\begin{align*}
d_{\mathrm{LE}}^2(g_t(U_1),g_t(U_{2:J+1}^{(\bm{w})}))
&= \|\log (g_t(U_1)) - \log (g_t(U_{2:J+1}^{(\bm{w})}))\|^2_F\\
&= \alpha_t^2\|\log (U_1) - \sum_{j=2}^{J+1}w_j\log (U_j)\|^2_F
= \alpha_t^2d_{\mathrm{LE}}^2(U_1,U_{2:J+1}^{(\bm{w})}).
\end{align*}
This yields $\bigcap_{t=1}^{T_0}\mathbb{W}_t^* = \bigcap_{t=T_0+1}^T\mathbb{W}_t^* = \mathbb{W}^*$ and hence Assumption 3.1 holds.

Fourth, we consider the space of SPD matrices with the Log-Cholesky metric. We identify each SPD matrix with its unique Cholesky factor $L\in\mathcal L_m^+$, where $\mathcal L_m^+$ is the space of lower-triangular matrices with positive diagonal entries. For $L\in\mathcal L_m^+$, let $\lfloor L\rfloor$ and $\mathbb D(L)$ denote its strictly lower-triangular and diagonal parts, respectively. The Log-Cholesky metric is defined as
\[
d_{\mathrm{LC}}(L_1,L_2) = \left\{\|\lfloor L_1 \rfloor - \lfloor L_2 \rfloor\|^2_F + \|\log (\mathbb{D}(L_1)) - \log (\mathbb{D}(L_2))\|^2_F\right\}^{1/2}.
\]
Note that the space $(\mathcal{L}^+_m, d_{\mathrm{LC}})$ is a geodesic metric space \citep{lin:19:1} and the geodesic with respect to $d_{\mathrm{LC}}$ between $L_1$ and $L_2$ is
\[
\gamma_{L_1,L_2}(t) = \lfloor L_1 \rfloor + t(\lfloor L_2 \rfloor - \lfloor L_1 \rfloor) + \exp\left\{\log (\mathbb{D}(L_1)) + t(\log (\mathbb{D}(L_2)) - \log (\mathbb{D}(L_1)))\right\}.
\]

Let $U_j \in \mathcal{L}^+_m$, $j=1,\dots,J+1$ and $\mu_t \in \mathcal{L}^+_m$, $t=1,\dots,T$. Recall that $g_t(U_j) = \gamma_{\mu_t,U_j}(\alpha_t)$. For $\bm{w} \in \Delta^{J-1}$,
\begin{align*}
\lfloor (U_{2:J+1}^{(\bm{w})}) \rfloor &= \sum_{j=2}^{J+1}w_j \lfloor U_j \rfloor,\ \log (\mathbb{D}(U_{2:J+1}^{(\bm{w})})) = \sum_{j=2}^{J+1}w_j\log (\mathbb{D}(U_j)),\\
\lfloor g_t^{(\bm{w})}(U_{2:J+1}) \rfloor
&= \lfloor \mu_t \rfloor + \alpha_t\left\{\sum_{j=2}^{J+1}w_j\lfloor U_j \rfloor - \lfloor \mu_t \rfloor \right\} = \lfloor g_t(U_{2:J+1}^{(\bm{w})}) \rfloor,\\
\log (\mathbb{D}(g_t^{(\bm{w})}(U_{2:J+1})))
&= \log (\mathbb{D}(\mu_t)) + \alpha_t\left\{\sum_{j=2}^{J+1}w_j\log (\mathbb{D}(U_j)) - \log (\mathbb{D}(\mu_t))\right\} \\
&= \log (\mathbb{D}(g_t(U_{2:J+1}^{(\bm{w})}))).
\end{align*}
Then
\begin{align*}
&d_{\mathrm{LC}}^2(g_t(U_1),g_t(U_{2:J+1}^{(\bm{w})}))
\\&= \|\lfloor g_t(U_1) \rfloor - \lfloor g_t(U_{2:J+1}^{(\bm{w})}) \rfloor \|^2_F + \|\log (\mathbb{D}(g_t(U_1))) - \log (\mathbb{D}(g_t(U_{2:J+1}^{(\bm{w})}))) \|^2_F\\
&= \alpha_t^2\|\lfloor U_1 \rfloor - \sum_{j=2}^{J+1}w_j\lfloor U_j \rfloor \|^2_F + \alpha_t^2\|\log (\mathbb{D}(U_1)) - \sum_{j=2}^{J+1}w_j\log (\mathbb{D}(U_j)) \|^2_F\\
&= \alpha_t^2d_{\mathrm{LC}}^2(U_1,U_{2:J+1}^{(\bm{w})}).
\end{align*}
This yields $\bigcap_{t=1}^{T_0}\mathbb{W}_t^* = \bigcap_{t=T_0+1}^T\mathbb{W}_t^* = \mathbb{W}^*$ and hence Assumption 3.1 holds.

\subsection{Univariate probability distributions }\label{exm:mea-app}
Recall that $g_t(U_j) := \gamma_{\mu_t, U_j}(\alpha_t)$, $\alpha_t \in (0,1)$. Then we have $F_{g_t(U_j)}^{-1}(\cdot) = F_{\mu_t}^{-1}(\cdot) + \alpha_t(F_{U_j}^{-1}(\cdot) - F_{\mu_t}^{-1}(\cdot))$. For $\bm{w} \in \Delta^{J-1}$,
\begin{align*}
F_{U_{2:J+1}^{(\bm{w})}}^{-1}(\cdot) &= \sum_{j=2}^{J+1}w_jF_{U_j}^{-1}(\cdot),\\
F_{g_t^{(\bm{w})}(U_{2:J+1})}^{-1}(\cdot) &= \sum_{j=2}^{J+1}w_jF_{g_t(U_j)}^{-1}(\cdot) = \sum_{j=2}^{J+1}w_j\left\{F_{\mu_t}^{-1}(\cdot) + \alpha_t(F_{U_j}^{-1}(\cdot) - F_{\mu_t}^{-1}(\cdot))\right\}\\
&= F_{\mu_t}^{-1}(\cdot) + \alpha_t\left\{\sum_{j=2}^{J+1}w_jF_{U_j}^{-1}(\cdot) - F_{\mu_t}^{-1}(\cdot)\right\} = \left(F_{\gamma_{\mu_t, U_{2:J+1}^{(\bm{w})}}(\alpha_t)}\right)^{-1}(\cdot)\\
&= F_{g_t(U_{2:J+1}^{(\bm{w})})}^{-1}(\cdot).
\end{align*}
Furthermore,
\begin{align*}
d_{\mathcal{W}}^2(g_t(U_1),g_t(U_{2:J+1}^{(\bm{w})})) &= \alpha_t^2\int_0^1\left\{F_{U_1}^{-1}(p) - \sum_{j=2}^{J+1}w_jF_{U_j}^{-1}(p)\right\}^2dp= \alpha_t^2d_{\mathcal{W}}^2(U_1, U_{2:J+1}^{(\bm{w})}),
\end{align*}
which yields $\bigcap_{t=1}^{T_0}\mathbb{W}_t^* = \bigcap_{t=T_0+1}^T\mathbb{W}_t^* = \mathbb{W}^*$ and hence Assumption 3.1 holds.

\subsection{Functional data}\label{exm:fun-app}
Note that $(g_t(U_j))(s) = (\gamma_{\mu_t, U_j}(\alpha_t))(s)
= \mu_t(s) + \alpha_t(U_j(s) - \mu_t(s))$, $s \in \mathcal{T}$. For $\bm{w} \in \Delta^{J-1}$,
\begin{align*}
U_{2:J+1}^{(\bm{w})}(s)
&= \sum_{j=2}^{J+1}w_jU_j(s),\\
(g_t^{(\bm{w})}(U_{2:J+1}))(s)
&= \sum_{j=2}^{J+1}w_j(g_t(U_j))(s)\\
&= \mu_t(s) + \alpha_t(U_{2:J+1}^{(\bm{w})}(s) - \mu_t(s)) = (g_t(U_{2:J+1}^{(\bm{w})}))(s),\ s \in \mathcal{T}.
\end{align*}
Next,
\begin{align*}
d_{L^2}^2(g_t(U_1),g_t(U_{2:J+1}^{(\bm{w})}))
&= \int_{\mathcal{T}}\left\{(g_t(U_1))(s) - (g_t(U_{2:J+1}^{(\bm{w})}))(s)\right\}^2 ds \\
&= \alpha_t^2\int_{\mathcal{T}}\left\{U_1(s) - U_{2:J+1}^{(\bm{w})}(s)\right\}^2ds
= \alpha_t^2d_{L^2}^2(U_1, U_{2:J+1}^{(\bm{w})}),
\end{align*}
which gives  $\bigcap_{t=1}^{T_0}\mathbb{W}_t^* = \bigcap_{t=T_0+1}^T\mathbb{W}_t^* = \mathbb{W}^*$ and hence Assumption 3.1 holds.

\section{\texorpdfstring{Verification of Assumption 3.2 and perturbation maps}{Verification of Assumption 3.2 and perturbation maps}}\label{sec:supp-perturbation}
This section provides sufficient conditions for Assumption~3.2 and the standing bounded-range condition in the outcome spaces considered in the paper. We first give sufficient conditions for the criterion convergence, minimizer, and stability requirements in Assumption~3.2. We then construct outcome-specific perturbation maps and record conditions under which the general results apply; the range implications use the signal models in Section~S.2. We write $a\geq0$ for the perturbation scale. Theorem~3.1 permits a fixed perturbation scale, while Corollary~\ref{cor:GSCM-consist} uses a triangular array with $a=a_{T_0}\to0$.

\subsection{\texorpdfstring{General sufficient conditions for Assumption 3.2}{General sufficient conditions for Assumption 3.2}}

Assumption~3.2 is verified through two alternative results. Proposition~\ref{prp:verify-coordinate} is the Hilbert-coordinate route: cross-moment convergence establishes criterion convergence, and coordinate linearity gives weight stability with exponent $\kappa=1$. Proposition~\ref{prp:verify-spherical} is the spherical-composition route: a loss-process law of large numbers establishes criterion convergence, and quadratic growth gives weight stability with exponent $\kappa=1/2$. Neither proposition depends on the other. Section~S.3.2 provides temporal conditions for the convergence requirements, and Sections~S.3.3--S.3.7 assign each outcome space to the appropriate route.

For the Hilbert-coordinate route, suppose that $\Phi:\mathcal M\to\mathcal H$ is an isometry onto a convex subset of a separable Hilbert space $\mathcal H$. For $t\leq T_0$, write
\[
X_{j,t}=\Phi(Y_{j,t}(N)),\qquad j=1,\ldots,J+1.
\]
Convexity of $\Phi(\mathcal M)$ implies that the weighted Fr\'echet mean $m_t(\bm w)=\widehat g_{2:J+1,t}^{(\bm w)}$ is unique and satisfies
\begin{equation}\label{eq:coordinate-mean}
\Phi(m_t(\bm w))=\sum_{j=2}^{J+1}w_jX_{j,t}.
\end{equation}
Fix $\nu_0\in\mathcal M$, write $z_0=\Phi(\nu_0)$, and define the empirical cross-moments
\[
G_{jk,T_0}=\frac{1}{T_0}\sum_{t=1}^{T_0}
\langle X_{j,t}-z_0,X_{k,t}-z_0\rangle_{\mathcal H},\qquad
1\leq j,k\leq J+1.
\]

\begin{proposition}[Hilbert-coordinate verification]\label{prp:verify-coordinate}
Suppose that $\|X_{j,t}-z_0\|_{\mathcal H}\leq B$ almost surely, uniformly over the required unit, time, and triangular-array indices, for some $B<\infty$. If
\[
G_{jk,T_0}\stackrel{p}{\longrightarrow}G_{jk}
\quad\text{for every }1\leq j,k\leq J+1,
\]
where the limits are deterministic, then Assumption~3.2 holds. Moreover, the pre- and post-treatment stability conditions hold with exponent $\kappa=1$ whenever the same coordinate bound holds over the corresponding time indices.
\end{proposition}

\begin{proof}
Since $\sum_{j=2}^{J+1}w_j=1$, the coordinate representation gives
\begin{align*}
\widehat Q(\bm w)
&=G_{11,T_0}-2\sum_{j=2}^{J+1}w_jG_{1j,T_0}
+\sum_{j,k=2}^{J+1}w_jw_kG_{jk,T_0}.
\end{align*}
There are finitely many cross-moments because $J$ is fixed. Their convergence therefore yields uniform convergence of $\widehat Q$ over $\Delta^{J-1}$ to
\[
\widetilde Q(\bm w)=G_{11}-2\sum_{j=2}^{J+1}w_jG_{1j}
+\sum_{j,k=2}^{J+1}w_jw_kG_{jk}.
\]
The cross-moments are uniformly bounded, so convergence in probability also implies convergence of their expectations. Thus this polynomial agrees with the definition $\widetilde Q(\bm w)=\lim_{T_0\to\infty}\E[\widehat Q(\bm w)]$. It is finite and continuous. Hence its full minimizer set $\widetilde{\mathcal W}$ is nonempty and compact, and continuity on the compact simplex gives the separation condition in Assumption~3.2(i). The empirical criterion is jointly measurable and continuous in $\bm w$; compactness of the simplex and the measurable selection theorem give a measurable empirical minimizer.

For any $\bm w_1,\bm w_2\in\Delta^{J-1}$, \eqref{eq:coordinate-mean} gives
\begin{align*}
d(m_t(\bm w_1),m_t(\bm w_2))
&=\left\|\sum_{j=2}^{J+1}(w_{1j}-w_{2j})(X_{j,t}-z_0)\right\|_{\mathcal H}\\
&\leq B\|\bm w_1-\bm w_2\|_1.
\end{align*}
This bound verifies Assumption~3.2(ii) and, when imposed over a fixed post-treatment horizon, the stability condition in Theorem~3.1(i).
\end{proof}

For the spherical-composition route, let $K$ be a common compact geodesically convex subset of the interior of $\mathcal S_+^{d-1}$ containing the relevant untreated outcomes and all of their weighted Fr\'echet means. For
\[
F_{t,\bm w}(\nu)=\sum_{j=2}^{J+1}w_jd^2(\nu,Y_{j,t}(N)),
\]
suppose that the weighted Fr\'echet mean $m_t(\bm w)$ is unique and jointly measurable and that, for some $c>0$,
\begin{equation}\label{eq:spherical-growth}
F_{t,\bm w}(\nu)-F_{t,\bm w}(m_t(\bm w))
\geq c\,d^2(\nu,m_t(\bm w))
\end{equation}
uniformly in $t$, $\bm w$, and $\nu\in K$. This quadratic-growth condition follows, for example, from uniform strong geodesic convexity of the squared-distance objectives on a sufficiently small spherical region.

\begin{proposition}[Spherical-composition verification]\label{prp:verify-spherical}
Suppose the preceding conditions hold and, for each $\bm w\in\Delta^{J-1}$,
\begin{align}
&\frac{1}{T_0}\sum_{t=1}^{T_0}\E[\ell_t(\bm w)]\longrightarrow\widetilde Q(\bm w),
\qquad
\ell_t(\bm w)=d^2(Y_{1,t}(N),m_t(\bm w)), \label{eq:loss-cesaro}\\
&\frac{1}{T_0}\sum_{t=1}^{T_0}
\{\ell_t(\bm w)-\E[\ell_t(\bm w)]\}\stackrel{p}{\longrightarrow}0. \label{eq:loss-lln}
\end{align}
Then Assumption~3.2 holds. The corresponding stability condition holds with exponent $\kappa=1/2$.
\end{proposition}

\begin{proof}
Let $D_K=\mathrm{diam}(K)$. Applying \eqref{eq:spherical-growth} with $\nu=m_t(\bm w_2)$ and comparing the two weighted objectives yields
\[
c\,d^2(m_t(\bm w_1),m_t(\bm w_2))
\leq 2D_K^2\|\bm w_1-\bm w_2\|_1.
\]
Thus $m_t(\bm w)$ is uniformly H\"older continuous in the weights with exponent $1/2$, which verifies Assumption~3.2(ii). Equations~\eqref{eq:loss-cesaro}--\eqref{eq:loss-lln} give pointwise convergence of $\widehat Q$. The same H\"older bound and the boundedness of $K$ imply that $\widetilde Q$ is continuous. Compactness then gives a nonempty minimizer set and separation, while joint measurability and continuity give a measurable empirical minimizer, as in the proof of Proposition~\ref{prp:verify-coordinate}.
\end{proof}

\subsection{\texorpdfstring{Temporal conditions for criterion convergence}{Temporal conditions for criterion convergence}}

The cross-moment condition in Proposition~\ref{prp:verify-coordinate} and the loss law of large numbers in Proposition~\ref{prp:verify-spherical} can be verified under standard temporal conditions. If the joint coordinate-outcome process is strictly stationary and ergodic, the ergodic theorem gives convergence of the bounded cross-moments. Likewise, if $\{\ell_t(\bm w)\}_{t\in\mathbb Z}$ is strictly stationary and ergodic for each fixed $\bm w$, then \eqref{eq:loss-cesaro}--\eqref{eq:loss-lln} hold with $\widetilde Q(\bm w)=\E[\ell_0(\bm w)]$.

Strict stationarity alone is not sufficient for either conclusion. A second sufficient regime, which also accommodates deterministic time variation in $g_t$, is that the joint innovation process is strongly mixing with coefficients $\alpha(k)\to0$, the relevant cross-products or losses are measurable functions of the current innovation vector, and their Ces\`aro expectations converge. Because these variables are uniformly bounded, the standard mixing covariance inequality gives
\[
\Var\left[\frac{1}{T_0}\sum_{t=1}^{T_0}\{Z_t-\E(Z_t)\}\right]
\leq \frac{C_1}{T_0}+\frac{C_2}{T_0}\sum_{k=1}^{T_0-1}\alpha(k)\longrightarrow0,
\]
where $Z_t$ denotes any required cross-product or loss. This proves the corresponding law of large numbers by Chebyshev's inequality. When $g_t$ varies deterministically, stationarity or weak dependence is imposed on the induced coordinate or loss process in this manner.

For the perturbation constructions below, let $\mathcal P_t=(\mathcal P_{1,t},\ldots,\mathcal P_{J+1,t})$. The sequence $\{\mathcal P_t\}_{t\in\mathbb Z}$ is strictly stationary if, for every positive integer $k$ and all integers $t_1,\ldots,t_k,\tau$, the joint distributions of $(\mathcal P_{t_1},\ldots,\mathcal P_{t_k})$ and $(\mathcal P_{t_1+\tau},\ldots,\mathcal P_{t_k+\tau})$ are identical. The stationary-ergodic regime is imposed on the induced coordinate-outcome or loss process, not only on the innovations. In the mixing regime, the full innovation vector across units is jointly strongly mixing, and current-time measurable transformations inherit its mixing coefficients. For a triangular array with row-specific coefficients $\alpha_{T_0}(k)$, the sufficient uniform condition is
\[
\frac{1}{T_0}\sum_{k=1}^{T_0-1}\alpha_{T_0}(k)\longrightarrow0,
\]
together with the uniform boundedness and Ces\`aro expectation convergence stated above.

\subsection{Networks}\label{subsec:supp-perturb-net}
Consider weighted undirected networks on $m$ nodes, represented by graph Laplacians. Let $\mathcal{E}_m=\{(r,s):1\le r<s\le m\}$ be the set of unordered edges and let $B_{rs}=(e_r-e_s)(e_r-e_s)^\top$, where $e_r$ is the $r$th Euclidean basis vector in $\mathbb{R}^m$. Any weighted graph Laplacian with nonnegative edge weights $b=(b_{rs})_{(r,s)\in\mathcal{E}_m}$ can be written as
\[
L(b)=\sum_{(r,s)\in\mathcal{E}_m}b_{rs}B_{rs}.
\]
Let $\xi_t=(\xi_{j,t,rs}:j=1,\ldots,J+1,(r,s)\in\mathcal E_m)$ be a joint innovation vector. Assume that the induced coordinate-outcome process is stationary and ergodic, or that $\xi_t$ and the cross-moment expectations satisfy the mixing regime above. Define
\[
\mathcal{P}_{j,t}(L(b))=L(b^{(j,t)}),\qquad
b_{rs}^{(j,t)}=b_{rs}\exp(a\xi_{j,t,rs}),\quad (r,s)\in\mathcal{E}_m.
\]
This map preserves nonnegative edge weights and the support of the network. If the latent and reference edge weights and $|\xi_{j,t,rs}|$ are uniformly bounded and $a$ ranges over a bounded interval, then
$\|\mathcal P_{j,t}(L(b))-L(b)\|_F\leq C a$ uniformly in $j$ and $t$, for a finite constant $C$. With $\Phi(L)=L$, graph Laplacians form a convex subset of the Hilbert space of symmetric matrices. Thus the bounded-range and stability conditions hold, and Proposition~\ref{prp:verify-coordinate} verifies Assumption~3.2 whenever the empirical matrix cross-moments satisfy its temporal condition.

\subsection{Compositional data}\label{subsec:supp-perturb-comp}

For the spherical representation based on the square-root transformation, let $\bm z=\sqrt{\bm y}\in\mathcal S_+^{d-1}$. Let $\zeta_t=(\zeta_{1,t},\ldots,\zeta_{J+1,t})$ be a joint innovation vector, with $\zeta_{j,t}\in\mathbb R^d$. Assume that the induced loss process is stationary and ergodic, or that $\zeta_t$ and the loss expectations satisfy the mixing regime above, and define
$v_{j,t}(\bm z)=(\bm I_d-\bm z\bm z^\top)\zeta_{j,t}\in T_{\bm z}\mathcal S^{d-1}$. The corresponding perturbation is
\[
\mathcal P_{j,t}(\bm z)=\mathrm{Exp}_{\bm z}(a v_{j,t}(\bm z)).
\]
Assume that the latent objects satisfy $\inf_{1\leq j\leq J+1}d(U_j,\partial\mathcal S_+^{d-1})\geq\eta$ for some $\eta>0$, that $\|\zeta_{j,t}\|\leq M$ almost surely, and that $aM<\eta$. Since $\|v_{j,t}(U_j)\|\leq M$, these conditions ensure that $\mathcal P_{j,t}(U_j)\in\mathcal S_+^{d-1}$ almost surely. Assume that there are a center $z_c\in\mathcal S_+^{d-1}$ and radii $0<\rho<r<\pi/4$ such that all unperturbed outcomes $g_t(U_j)$ lie in the closed spherical ball of radius $\rho$ centered at $z_c$. Suppose also that the maps $g_t$ are uniformly Lipschitz on the relevant range, $\|\zeta_{j,t}\|$ is uniformly bounded, and $a$ is small enough that all perturbed outcomes $g_t(\mathcal P_{j,t}(U_j))$ lie in the concentric closed ball $K$ of radius $r$. The perturbed compositions satisfy $d(\mathcal P_{j,t}(\bm z),\bm z)\leq Ca$, and their weighted Fr\'echet means remain in $K$. On $K$, the squared-distance objectives are uniformly strongly geodesically convex. Continuity of the weighted squared-distance objective, compactness of $K$, and uniqueness of its minimizer imply joint measurability of $m_t(\bm w)$ by the measurable maximum theorem. Proposition~\ref{prp:verify-spherical} therefore verifies Assumption~3.2 under either temporal regime above.

\subsection{SPD matrices}\label{subsec:supp-perturb-spd}
Let $\mathcal{M}=\mathrm{Sym}^+_m$ be the space of $m\times m$ SPD matrices. Under the Log-Euclidean metric, perturbations can be introduced additively after applying the matrix logarithm. Specifically, let $\Xi_t=(\Xi_{1,t},\ldots,\Xi_{J+1,t})$ be a joint innovation vector with symmetric-matrix components. Assume that the induced coordinate-outcome process is stationary and ergodic, or that $\Xi_t$ and the cross-moment expectations satisfy the mixing regime above, and define
\[
\mathcal P_{j,t}(A)=\exp\{\log A+a\Xi_{j,t}\},\qquad A\in\mathrm{Sym}^+_m.
\]
The map is well defined and
$d_{\mathrm{LE}}(\mathcal P_{j,t}(A),A)=a\|\Xi_{j,t}\|_F$.

More generally, suppose an SPD metric is represented as a Hilbert distance after an invertible transformation $\Phi$. Let $\Xi_{j,t}$ take values in the corresponding coordinate space, and assume that $\Phi(A)+a\Xi_{j,t}$ belongs to $\Phi(\mathrm{Sym}^+_m)$ almost surely. Then set
\[
\mathcal P_{j,t}(A)=\Phi^{-1}(\Phi(A)+a\Xi_{j,t}).
\]
For the Frobenius, power, and Log-Euclidean metrics, one may take $\Phi(A)=A$, $A^p$, and $\log A$, respectively. For the Log-Cholesky metric, $\Phi$ consists of the strictly lower triangular Cholesky entries and logarithms of its diagonal entries. For the identity and power transformations, membership in the transformed SPD space follows, for example, from a uniform eigenvalue margin and a sufficiently small upper bound on $a$. If the transformed latent and reference matrices and $\Xi_{j,t}$ are uniformly bounded and $a$ ranges over such a bounded interval, the standing bounded-range condition holds and the perturbations have size $O(a)$ in the corresponding metric. Proposition~\ref{prp:verify-coordinate} then verifies Assumption~3.2 under its cross-moment condition.

\subsection{Univariate probability distributions}\label{subsec:supp-perturb-dist}
Consider univariate probability distributions on $\mathbb R$ with finite second moments, equipped with the 2-Wasserstein metric. Let $\xi_t=(\xi_{1,t},\ldots,\xi_{J+1,t})$ be a joint random-function innovation vector. Assume that the induced coordinate-outcome process is stationary and ergodic, or that $\xi_t$ and the cross-moment expectations satisfy the mixing regime above. Assume also that, almost surely, $T_{j,t,a}(x)=x+a\xi_{j,t}(x)$ is nondecreasing and Borel measurable and that $T_{j,t,a}(X)$ has a finite second moment whenever $X$ does. Define $\mathcal P_{j,t}(\mu)=(T_{j,t,a})_\#\mu$. Its quantile function is
\[
F_{\mathcal P_{j,t}(\mu)}^{-1}(s)=T_{j,t,a}(F_\mu^{-1}(s)),\qquad\text{for Lebesgue-almost every }s\in(0,1).
\]
Moreover,
\[
d_{\mathcal W}(\mathcal P_{j,t}(\mu),\mu)
\leq a\left[\int|\xi_{j,t}(x)|^2\,d\mu(x)\right]^{1/2}.
\]
Uniform bounds on the right-hand side therefore give perturbations of order $a$. A common bounded support for the latent and reference distributions, together with a uniform bound on $\xi_{j,t}$ over that support and bounded $a$, gives the standing bounded-range condition. The quantile map $\Phi(\mu)=F_\mu^{-1}$ embeds this space isometrically into a convex subset of $L^2(0,1)$, so Proposition~\ref{prp:verify-coordinate} verifies Assumption~3.2 under its cross-moment condition.

\subsection{Functional data}\label{subsec:supp-perturb-fun}
Let $\mathcal M=L^2(\mathcal T)$, where $\mathcal T$ is a compact interval, equipped with the usual $L^2$ metric. Let $\xi_t=(\xi_{1,t},\ldots,\xi_{J+1,t})$ be a joint $L^2(\mathcal T)$-valued innovation vector. Assume that the induced coordinate-outcome process is stationary and ergodic, or that $\xi_t$ and the cross-moment expectations satisfy the mixing regime above, and define
\[
\mathcal P_{j,t}(f)=f+a\xi_{j,t},\qquad f\in L^2(\mathcal T).
\]
Then $d_{L^2}(\mathcal P_{j,t}(f),f)=a\|\xi_{j,t}\|_{L^2}$. Uniformly bounded latent and reference $L^2$ norms, almost surely bounded innovation norms, and bounded $a$ give the standing bounded-range condition; uniformly bounded first moments suffice for the order-$a$ expectation bound used in Corollary~\ref{cor:GSCM-consist}. Taking $\Phi$ to be the identity, Proposition~\ref{prp:verify-coordinate} verifies Assumption~3.2 under its cross-moment condition.

\subsection{\texorpdfstring{Summary of the verification}{Summary of the verification}}

Propositions~\ref{prp:verify-coordinate} and~\ref{prp:verify-spherical} verify every component of Assumption~3.2: pointwise criterion convergence, continuity and existence of the possibly nonunique limiting minimizer set, separation from that set, existence of a measurable empirical minimizer, and the uniform local stability condition. The temporal assumptions are used specifically to establish the law of large numbers in Assumption~3.2(i); the geometric arguments establish Assumption~3.2(ii).

For completeness, the same constructions also provide primitive conditions for Assumption~\ref{ass:GSCM-consist}(ii). In the Hilbert-coordinate cases, for donor tuples $x=(x_2,\ldots,x_{J+1})$ and $x'=(x'_2,\ldots,x'_{J+1})$,
\[
d(m^{(\bm w)}(x),m^{(\bm w)}(x'))
\leq\max_{2\leq j\leq J+1}d(x_j,x'_j).
\]
If the maps $g_t$ are uniformly Lipschitz on the relevant ranges, the order-$a$ perturbation bounds therefore give Assumption~\ref{ass:GSCM-consist}(ii) with $\delta_{T_0}=a_{T_0}$. For spherical compositions, the same uniform Lipschitz condition on $g_t$, together with the quadratic-growth comparison, gives an order-$a_{T_0}^{1/2}$ bound, so one may take $\delta_{T_0}=a_{T_0}^{1/2}$; a uniform Lipschitz sensitivity bound for the Fr\'echet means gives the sharper order $a_{T_0}$. The latent balancing and persistence requirements used in Assumption~\ref{ass:GSCM-consist}(i) are treated in Section~S.2.

\begin{center}
\begin{minipage}{\textwidth}
\captionof{table}{Sufficient conditions verifying the standing bounded-range condition and Assumption~3.2. The temporal condition in the last column is imposed on the induced process or the full innovation vector across units, as specified in Section~S.3.2.}
\label{tab:assumption-verification}
{\single\small
\begin{tabular}{>{\raggedright\arraybackslash}p{.14\textwidth}>{\raggedright\arraybackslash}p{.34\textwidth}>{\raggedright\arraybackslash}p{.40\textwidth}}
\hline
Outcome & Range and geometry & Verification of Assumption~3.2\\
\hline
Networks & Uniformly bounded latent and reference edge weights and bounded edge innovations; Frobenius coordinates. & Proposition~\ref{prp:verify-coordinate}; stationary-ergodic or mixing convergence of the matrix cross-moments.\\[3pt]
Compositions & A common closed spherical ball of radius less than $\pi/4$ in the interior of the positive orthant; uniformly bounded tangent perturbations. & Proposition~\ref{prp:verify-spherical}; stationary-ergodic or mixing law of large numbers for the bounded loss process.\\[3pt]
SPD matrices & Uniformly bounded coordinates under the identity, power, logarithm, or Log-Cholesky transformation, as appropriate; bounded coordinate innovations that remain in the transformed SPD space. & Proposition~\ref{prp:verify-coordinate}; stationary-ergodic or mixing convergence of the transformed-matrix cross-moments.\\[3pt]
Distributions & Common bounded support, or uniformly bounded quantile functions in $L^2(0,1)$; perturbations remain in the corresponding bounded set. & Proposition~\ref{prp:verify-coordinate} with quantile coordinates; stationary-ergodic or mixing convergence of the quantile cross-moments.\\[3pt]
Functional data & Uniformly bounded latent and reference $L^2$ norms and almost surely bounded innovation norms. & Proposition~\ref{prp:verify-coordinate} with identity coordinates; stationary-ergodic or mixing convergence of the $L^2$ cross-moments.\\
\hline
\end{tabular}}
\end{minipage}
\end{center}

\section{Proofs}
\subsection{Proof of Theorem 3.1}

\begin{proof}[Proof of (i)]
For $t\leq T_0$, $Y_{j,t}=Y_{j,t}(N)$ for every unit, and write $m_t(\bm w)=\widehat g_{2:J+1,t}^{(\bm w)}$.
The standing bounded-range condition in Section~2 places the treated and donor outcomes in the closed ball of radius $D$ centered at $\nu_0$. By optimality of the weighted Fr\'echet mean,
\[
\sum_{j=2}^{J+1}w_jd^2(Y_{j,t}(N),m_t(\bm w))
\leq\sum_{j=2}^{J+1}w_jd^2(Y_{j,t}(N),\nu_0)\leq D^2.
\]
The triangle and Cauchy--Schwarz inequalities therefore give
\[
d(\nu_0,m_t(\bm w))
\leq D+\sum_{j=2}^{J+1}w_jd(Y_{j,t}(N),m_t(\bm w))\leq2D.
\]
Consequently, $d(Y_{1,t}(N),m_t(\bm w))\leq3D$ for every $\bm w\in\Delta^{J-1}$. For any $\bm w_1,\bm w_2\in\Delta^{J-1}$,
\begin{align}
|\widehat Q(\bm w_1)-\widehat Q(\bm w_2)|
&\leq \frac{6D}{T_0}\sum_{t=1}^{T_0}
d(m_t(\bm w_1),m_t(\bm w_2)). \label{eq:criterion-modulus}
\end{align}
Assumption~3.2(ii) and \eqref{eq:criterion-modulus} imply stochastic equicontinuity of $\widehat Q$ on $\Delta^{J-1}$.

Pointwise convergence in Assumption~3.2(i) gives finite-dimensional convergence of $\widehat Q$ to the deterministic process $\widetilde Q$ and pointwise asymptotic tightness. Moreover, the simplex is compact and hence totally bounded under the $\ell_1$ metric, and continuity of $\widetilde Q$ implies uniform continuity. Stochastic equicontinuity and Theorem~1.5.7 of \citet{well:96} give asymptotic tightness of $\widehat Q$. Since the deterministic continuous process $\widetilde Q$ is a tight element of $\ell^\infty(\Delta^{J-1})$, finite-dimensional convergence and Theorem~1.5.4 of \citet{well:96} then give
\[
\widehat Q\rightsquigarrow\widetilde Q
\quad\text{in }\ell^\infty(\Delta^{J-1}).
\]
Applying the continuous mapping theorem (Theorem~1.3.6 of \citet{well:96}) to the continuous map $f\mapsto\|f-\widetilde Q\|_\infty$, and using that the limit is deterministic, yields
\begin{equation}\label{lim:GSCM-w-conv}
\sup_{\bm w\in\Delta^{J-1}}|\widehat Q(\bm w)-\widetilde Q(\bm w)|=o_p(1).
\end{equation}

We next adapt the M-estimator consistency argument in Corollary~3.2.3 of \citet{well:96} to the possibly non-singleton minimizer set. Fix $\varepsilon>0$. If every $\bm w\in\Delta^{J-1}$ satisfies $d_1(\bm w,\widetilde{\mathcal W})<\varepsilon$, the desired conclusion is immediate. Otherwise, the separation condition in Assumption~3.2(i) gives
\[
c_\varepsilon=
\inf_{\{\bm w\in\Delta^{J-1}:d_1(\bm w,\widetilde{\mathcal W})\geq\varepsilon\}}
\widetilde Q(\bm w)-\min_{\bm w\in\Delta^{J-1}}\widetilde Q(\bm w)>0.
\]
Let $q_0=\min_{\bm w\in\Delta^{J-1}}\widetilde Q(\bm w)$ and fix $\bm w_0\in\widetilde{\mathcal W}$. On the event that the supremum in \eqref{lim:GSCM-w-conv} is smaller than $c_\varepsilon/3$, any $\bm w$ with $d_1(\bm w,\widetilde{\mathcal W})\geq\varepsilon$ satisfies $\widehat Q(\bm w)>q_0+2c_\varepsilon/3$, whereas $\widehat Q(\bm w_0)<q_0+c_\varepsilon/3$. An empirical minimizer therefore cannot be at distance at least $\varepsilon$ from $\widetilde{\mathcal W}$. Hence
$d_1(\widehat{\bm w},\widetilde{\mathcal W})=o_p(1)$.

The nearest-point correspondence onto the deterministic compact set $\widetilde{\mathcal W}$ has nonempty compact values and a measurable selection. Let $\widetilde{\bm w}$ be the selection in the theorem, and let $L_{T_0}=O_p(1)$ denote the supremum in the post-treatment stability condition. At $t=T_0+h$,
\[
d(\widehat g_{2:J+1,t}^{(\widehat{\bm w})},
\widehat g_{2:J+1,t}^{(\widetilde{\bm w})})
\leq L_{T_0}\|\widehat{\bm w}-\widetilde{\bm w}\|_1^\kappa=o_p(1).
\]
The triangle inequality now gives the error bound stated in Theorem~3.1(i).
\end{proof}

\begin{proof}[Proof of (ii)]
For all sufficiently large $T_0$, $\widetilde{\bm w}$ belongs to
$\bigcap_{s=1}^{T_0}\mathbb W_s^*$ by the additional inclusion in part~(ii). Assumption~3.1(ii) then implies
$\widetilde{\bm w}\in\mathbb W_{T_0+h}^*$ for every fixed $h\geq1$, which is exactly the stated equality.
\end{proof}

We next give sufficient conditions for consistent recovery of the untreated counterfactual. Let $\{\mathcal P_{j,t}^{(T_0)}\}$ be a triangular array of perturbation maps and write
\[
Y_{j,t}^{(T_0)}(N)=g_t(\mathcal P_{j,t}^{(T_0)}(U_j)),
\qquad
\widehat g_{2:J+1,t,T_0}^{(\bm w)}
=\argmin_{\nu\in\mathcal M}\sum_{j=2}^{J+1}w_j
d^2(g_t(\mathcal P_{j,t}^{(T_0)}(U_j)),\nu).
\]
Define the row-$T_0$ empirical criterion and GSC weights by
\[
\widehat Q_{T_0}(\bm w)
=\frac{1}{T_0}\sum_{s=1}^{T_0}
d^2(Y_{1,s}^{(T_0)}(N),\widehat g_{2:J+1,s,T_0}^{(\bm w)}),
\qquad
\widehat{\bm w}_{T_0}\in\argmin_{\bm w\in\Delta^{J-1}}\widehat Q_{T_0}(\bm w).
\]
For $t>T_0$, define the row-$T_0$ GSC prediction by
\[
\widehat Y_{1,t,T_0}^{(\mathrm{GSC})}
=\widehat g_{2:J+1,t,T_0}^{(\widehat{\bm w}_{T_0})}.
\]
All notation in Theorem~3.1 is understood row by row for this array. Define the latent finite-sample criterion and its limit by
\[
Q_{T_0}^*(\bm w)=\frac{1}{T_0}\sum_{t=1}^{T_0}
d^2(g_t(U_1),g_{2:J+1,t}^{(\bm w)}),
\qquad
Q^*(\bm w)=\lim_{T_0\to\infty}Q_{T_0}^*(\bm w),
\]
and let $\mathcal W^*=\argmin_{\bm w\in\Delta^{J-1}}Q^*(\bm w)$.

\begin{assumption}\label{ass:GSCM-consist}\quad
\begin{itemize}
\item[(i)] The function $Q^*$ exists as a finite continuous function, $\mathcal W^*$ is nonempty, and every limiting oracle weight exactly balances the latent pre-treatment outcomes for all sufficiently large $T_0$:
\[
\mathcal W^*\subset\bigcap_{t=1}^{T_0}\mathbb W_t^*.
\]
\item[(ii)] There is a sequence $\delta_{T_0}>0$ with $\delta_{T_0}\to0$ such that, for every fixed $H\geq1$,
\begin{align*}
&\max_{1\leq t\leq T_0+H}
\E\left[d(g_t(\mathcal P_{1,t}^{(T_0)}(U_1)),g_t(U_1))\right]
=O(\delta_{T_0}),\\
&\max_{1\leq t\leq T_0+H}
\E\left[
\sup_{\bm w\in\Delta^{J-1}}
d(\widehat g_{2:J+1,t,T_0}^{(\bm w)},g_{2:J+1,t}^{(\bm w)})
\right]
=O(\delta_{T_0}).
\end{align*}
\end{itemize}
\end{assumption}

\begin{corollary}\label{cor:GSCM-consist}
Suppose that Assumptions~3.1, 3.2, and~\ref{ass:GSCM-consist} hold for the triangular array above. For a fixed $h\geq1$, suppose also that, for some $\kappa\in(0,1]$,
\[
\sup_{\substack{\bm w_1,\bm w_2\in\Delta^{J-1}\\\bm w_1\neq\bm w_2}}
\frac{d(\widehat g_{2:J+1,T_0+h,T_0}^{(\bm w_1)},
\widehat g_{2:J+1,T_0+h,T_0}^{(\bm w_2)})}
{\|\bm w_1-\bm w_2\|_1^\kappa}=O_p(1).
\]
Then
\[
d(Y_{1,T_0+h}^{(T_0)}(N),\widehat Y_{1,T_0+h,T_0}^{(\mathrm{GSC})})=o_p(1).
\]
No uniqueness of the oracle weights in $\mathcal W^*$ is required.
\end{corollary}

\begin{proof}
Let $\widehat Q_{T_0}$ denote the empirical perturbed criterion in row $T_0$. Under the standing bounded-range condition in Section~2, the argument used in the proof of Theorem~3.1 places every latent and perturbed synthetic object within distance $2D$ of $\nu_0$. Hence, uniformly over $\bm w\in\Delta^{J-1}$,
\begin{align*}
|\E[\widehat Q_{T_0}(\bm w)]-Q_{T_0}^*(\bm w)|
&\leq \frac{6D}{T_0}\sum_{t=1}^{T_0}
\E\left[d(g_t(\mathcal P_{1,t}^{(T_0)}(U_1)),g_t(U_1))\right.\\
&\hspace{42mm}\left.{}+
\sup_{\bm v\in\Delta^{J-1}}
d(\widehat g_{2:J+1,t,T_0}^{(\bm v)},g_{2:J+1,t}^{(\bm v)})
\right]\\
&=O(\delta_{T_0}).
\end{align*}
Taking limits gives $\widetilde Q(\bm w)=Q^*(\bm w)$ for every $\bm w$, and therefore
$\widetilde{\mathcal W}=\mathcal W^*$. Assumption~\ref{ass:GSCM-consist}(i), together with Assumption~3.1(ii), satisfies the conditions of Theorem~3.1(ii), so the latent-signal term in the bound in Theorem~3.1(i) is zero at $t=T_0+h$.

Let $\widetilde{\bm w}\in\mathcal W^*$ be the nearest-point selection used in Theorem~3.1. Markov's inequality and Assumption~\ref{ass:GSCM-consist}(ii) give
\begin{align*}
d(g_t(\mathcal P_{1,t}^{(T_0)}(U_1)),g_t(U_1))&=O_p(\delta_{T_0}),\\
d(\widehat g_{2:J+1,t,T_0}^{(\widetilde{\bm w})},
g_{2:J+1,t}^{(\widetilde{\bm w})})&=O_p(\delta_{T_0})
\end{align*}
at $t=T_0+h$. Applying the error bound in Theorem~3.1(i) completes the proof because $\delta_{T_0}\to0$.
\end{proof}

\subsection{\texorpdfstring{Proof of Proposition 4.1}{Proof of Proposition 4.1}}

Condition on the training-data $\sigma$-field $\mathcal F_{\mathrm{tr}}$, so that $\widehat{\bm w}^{\mathrm{tr}}$ is fixed. For $z=(z_1,\ldots,z_{J+1})\in\mathcal M^{J+1}$, let
\[
m(z_2,\ldots,z_{J+1})
=\argmin_{\nu\in\mathcal M}\sum_{j=2}^{J+1}\widehat w_j^{\mathrm{tr}}d^2(\nu,g(z_j)),
\qquad
r(z)=d(g(z_1),m(z_2,\ldots,z_{J+1})).
\]
By the assumed uniqueness and measurability of the weighted Fr\'echet mean map, $r$ is measurable. If $g_s=g$, then $R_s=r(Z_s)$. If $g_s=h_s\circ g$, isometric equivariance and uniqueness of the weighted Fr\'echet mean give
\[
\widehat Y_{1,s}^{(\mathrm{GSC,tr})}
=h_s(m(Z_{2,s},\ldots,Z_{J+1,s})).
\]
Therefore
\[
R_s=d\!\left(h_s(g(Z_{1,s})),h_s(m(Z_{2,s},\ldots,Z_{J+1,s}))\right)=r(Z_s).
\]
In either case, every score is the same measurable scalar function of the corresponding perturbation vector. Conditional exchangeability of the $Z_s$ therefore implies conditional exchangeability of the scores, which gives Assumption~4.1. No independence among the components of $Z_s$ is used.

For the global procedure, each block score is the same measurable function of the corresponding equal-length block of perturbation vectors. Exchangeability of these block vectors therefore implies exchangeability of the block scores in Assumption~4.2. Conditional i.i.d.\ sampling of the individual $Z_s$ over time is a sufficient special case for both conclusions.

\subsection{Proof of Theorem 4.1}
For each $s\in\mathcal T_{\mathrm{cal}}$, since $s\le T_0$, the treated unit is untreated during the calibration period and hence $Y_{1,s}=Y_{1,s}(N)$. Therefore the calibration scores satisfy
\[
R_s=d\left(Y_{1,s},\widehat Y_{1,s}^{(\mathrm{GSC,tr})}\right)=d\left(Y_{1,s}(N),\widehat Y_{1,s}^{(\mathrm{GSC,tr})}\right).
\]
By Assumption~4.1, conditional on the data used to estimate the GSC training weights, the collection $\{R_s:s\in\mathcal T_{\mathrm{cal}}\}\cup\{R_t\}$ is exchangeable, where $R_t=d\left(Y_{1,t}(N),\widehat Y_{1,t}^{(\mathrm{GSC,tr})}\right)$.

If $q_{1-\alpha}=+\infty$, then $\mathcal C_t^{(\mathrm{GSC})}(1-\alpha)=\mathcal M$, and the coverage claim is immediate. Hence it remains to consider the case $q_{1-\alpha}=R_{(k_\alpha)}$ for $k_\alpha\leq m$. By definition of the prediction region,
\[
Y_{1,t}(N)\notin \mathcal C_t^{(\mathrm{GSC})}(1-\alpha)
\]
if and only if
\[
d\left(Y_{1,t}(N),\widehat Y_{1,t}^{(\mathrm{GSC,tr})}\right)>q_{1-\alpha},
\]
or equivalently, $R_t>R_{(k_\alpha)}$. Since the $m+1$ scores are exchangeable, the rank of $R_t$ among $\{R_s:s\in\mathcal T_{\mathrm{cal}}\}\cup\{R_t\}$ is uniformly distributed over $\{1,\ldots,m+1\}$ in the absence of ties. In the presence of ties, the use of the strict inequality in the event $\{R_t>R_{(k_\alpha)}\}$ yields conservative coverage by the standard conformal prediction argument. Consequently,
\[
\Pr\{R_t>R_{(k_\alpha)}\}\le\frac{m+1-k_\alpha}{m+1}\le\alpha.
\]
Therefore,
\[
\Pr\left\{Y_{1,t}(N)\in \mathcal C_t^{(\mathrm{GSC})}(1-\alpha)\right\}\ge 1-\alpha.
\]

Under the pointwise null hypothesis
\[
H_{0,t}:Y_{1,t}(I)=Y_{1,t}(N),
\]
we have
\[
Y_{1,t}=Y_{1,t}(I)=Y_{1,t}(N).
\]
Hence
\[
\Pr_{H_{0,t}}\{\phi_t=1\}=\Pr_{H_{0,t}}\left\{Y_{1,t}(N)\notin \mathcal C_t^{(\mathrm{GSC})}(1-\alpha)\right\}\le \alpha,
\]
which establishes the finite-sample validity of the test.

\subsection{Proof of Corollary 4.1}
Since $(\mathcal M,d)$ is uniquely geodesic, for each $\omega\in\mathcal M$ there is a
unique geodesic $\gamma_{\omega,Y_{1,t}}$. On the event
\[
Y_{1,t}(N)\in \mathcal C_t^{(\mathrm{GSC})}(1-\alpha),
\]
we have
\[
\tau_t=\gamma_{Y_{1,t}(N),Y_{1,t}(I)}=\gamma_{Y_{1,t}(N),Y_{1,t}}\in\left\{\gamma_{\omega,Y_{1,t}}:\omega\in \mathcal C_t^{(\mathrm{GSC})}(1-\alpha)\right\}=\mathcal G_t(1-\alpha),
\]
where we used $Y_{1,t}=Y_{1,t}(I)$ for $t>T_0$. Therefore
\[
\Pr\{\tau_t\in \mathcal G_t(1-\alpha)\}\ge\Pr\{Y_{1,t}(N)\in \mathcal C_t^{(\mathrm{GSC})}(1-\alpha)\}\ge1-\alpha,
\]
by Theorem~4.1. This proves the result.

\subsection{Proof of Corollary 4.2}
Since $Y_{1,t}=Y_{1,t}(I)$ for $t>T_0$,
\[
\widehat\delta_t=d(\widehat Y^{(\mathrm{GSC,tr})}_{1,t},Y_{1,t}(I)).
\]
By the reverse triangle inequality,
\[
|\widehat\delta_t-\delta_t|=\left|d(\widehat Y^{(\mathrm{GSC,tr})}_{1,t},Y_{1,t}(I))-d(Y_{1,t}(N),Y_{1,t}(I))\right|\le d(\widehat Y^{(\mathrm{GSC,tr})}_{1,t},Y_{1,t}(N)).
\]
On the event $Y_{1,t}(N)\in\mathcal C_t^{(\mathrm{GSC})}(1-\alpha)$,
\[
d(\widehat Y^{(\mathrm{GSC,tr})}_{1,t},Y_{1,t}(N))\le q_{1-\alpha}.
\]
Thus, on this event,
\[
\delta_t\in[\widehat\delta_t-q_{1-\alpha},\widehat\delta_t+q_{1-\alpha}].
\]
Since $\delta_t\ge 0$, the lower endpoint can be replaced by $\{\widehat\delta_t-q_{1-\alpha}\}_+$. The coverage statement follows from Theorem~4.1.

\subsection{Proof of Theorem 4.2}
Under $H_0^{\mathrm{glob}}$, Assumption~4.2 implies that, conditional on the data used to estimate the GSC weights, the block scores $S_r(\mathcal B_1),\ldots,S_r(\mathcal B_K),S_r(\mathcal B_*)$ are exchangeable.

By definition, $\widehat p_r$ is the conformal rank $p$-value associated with the post-treatment block score $S_r(\mathcal B_*)$ relative to the calibration block scores $S_r(\mathcal B_1),\ldots,S_r(\mathcal B_K)$.

If there are no ties among the $K+1$ block scores, exchangeability implies that the rank of $S_r(\mathcal B_*)$ among $S_r(\mathcal B_1),\ldots,S_r(\mathcal B_K),S_r(\mathcal B_*)$ is uniformly distributed on $\{1,\ldots,K+1\}$. Consequently,
\[
\Pr_{H_0^{\mathrm{glob}}}\left\{\widehat p_r\le\alpha\right\}\le\alpha.
\]

In the presence of ties, the use of the non-strict comparison $S_r(\mathcal B_k)\ge S_r(\mathcal B_*)$ in the definition of $\widehat p_r$ yields a conservative conformal $p$-value. Therefore the same inequality continues to hold:
\[
\Pr_{H_0^{\mathrm{glob}}}\left\{\widehat p_r\le\alpha\right\}\le\alpha.
\]

Since
\[
\phi_r^{\mathrm{glob}}=1\{\widehat p_r\le\alpha\},
\]
it follows that
\[
\Pr_{H_0^{\mathrm{glob}}}\left\{\phi_r^{\mathrm{glob}}=1\right\}=\Pr_{H_0^{\mathrm{glob}}}\left\{\widehat p_r\le\alpha\right\}\le\alpha.
\]
Hence the global test has finite-sample size at most $\alpha$.

\section{\texorpdfstring{Additional simulation and application results}{Additional simulation and application results}}\label{app:additional-simulations}

We report the SPD counterpart to the network simulation in Section~5.2 of the main paper. In Log-Euclidean coordinates, the donor latent objects are $10\times10$ symmetric matrices with bounded entries, the treated latent object is their convex combination with oracle weights $\bm w^*$, and the reference object contains an analogous periodic common component. Independent perturbation matrices are formed by symmetrizing matrices of standardized $\mathrm{Unif}[-\sqrt{3},\sqrt{3}]$ entries and are added in log coordinates at the fixed scale $\sigma=.10$. Exponentiation maps the resulting symmetric matrices to SPD outcomes. All other estimation settings match the network experiment.

\begin{table}[H]
\single
\centering
\begin{tabular}{lccc}
\hline
Quantity & $T_0=20$ & $T_0=100$ & $T_0=500$ \\
\hline
Post-treatment prediction gap & .048 ($<$.001) & .022 ($<$.001) & .010 ($<$.001) \\
Realized post-treatment error & .387 (.001) & .386 (.001) & .386 (.001) \\
$\|\widehat{\bm w}-\widetilde{\bm w}\|_1$ & .174 (.002) & .081 (.001) & .036 ($<$.001) \\
$\|\widetilde{\bm w}-\bm w^*\|_1$ & .115 (.001) & .118 (.001) & .115 (.001) \\
\hline
\end{tabular}
\caption{SPD estimation results for $T_0\in\{20,100,500\}$ and $\sigma=.10$, based on 500 replications. The post-treatment prediction gap compares the post-treatment predictions based on $\widehat{\bm w}$ and $\widetilde{\bm w}$, while the realized post-treatment error compares the untreated post-treatment outcome with the prediction based on $\widehat{\bm w}$. The remaining rows report the $\ell_1$ weight-estimation error and oracle-weight gap. Entries are Monte Carlo means with Monte Carlo standard errors in parentheses.}
\label{tab:sim-spd}
\end{table}

Table~\ref{tab:sim-spd} shows that the post-treatment prediction gap and weight-estimation error decrease with $T_0$, whereas the realized post-treatment error and oracle-weight gap remain stable, consistent with the decomposition in Theorem~3.1. This pattern agrees with the network results.

To separate the effects of pre-treatment length and perturbation magnitude, we repeat the estimation experiment over $T_0\in\{20,100,500\}$ and $\sigma\in\{20^{-1/2},100^{-1/2},500^{-1/2}\}$. For each perturbation scale $\sigma$, let $\widetilde{\bm w}_\sigma$ denote the selected minimizer of $\widetilde Q$ under the corresponding simulation design. The latent objects and standardized innovations are paired across perturbation scales within each outcome, $T_0$, and replication. Table~\ref{tab:sim-perturbation-sensitivity} reports the four diagnostics used in the fixed-$\sigma$ experiments.

\begin{table}[H]
\single
\centering
\setlength{\tabcolsep}{4.5pt}
\begin{tabular}{llccc}
\hline
Outcome & $T_0$ & $\sigma=20^{-1/2}$ & $\sigma=100^{-1/2}$ & $\sigma=500^{-1/2}$ \\
\hline
\multicolumn{5}{l}{\textit{Panel A: Post-treatment prediction gap}} \\
Network & 20  & \textbf{.161 (.002)} & .068 (.001) & .030 ($<$.001) \\
              & 100 & .076 (.001) & \textbf{.031 ($<$.001)} & .014 ($<$.001) \\
              & 500 & .034 ($<$.001) & .014 ($<$.001) & \textbf{.006 ($<$.001)} \\
SPD     & 20  & \textbf{.107 (.001)} & .048 ($<$.001) & .022 ($<$.001) \\
              & 100 & .050 ($<$.001) & \textbf{.022 ($<$.001)} & .010 ($<$.001) \\
              & 500 & .022 ($<$.001) & .010 ($<$.001) & \textbf{.005 ($<$.001)} \\
\multicolumn{5}{l}{\textit{Panel B: Realized post-treatment error}} \\
Network & 20  & \textbf{.846 (.006)} & .374 (.003) & .167 (.001) \\
              & 100 & .844 (.006) & \textbf{.373 (.003)} & .167 (.001) \\
              & 500 & .830 (.006) & .367 (.003) & \textbf{.164 (.001)} \\
SPD     & 20  & \textbf{.861 (.003)} & .387 (.001) & .174 (.001) \\
              & 100 & .859 (.003) & \textbf{.386 (.001)} & .173 (.001) \\
              & 500 & .859 (.003) & .386 (.001) & \textbf{.173 (.001)} \\
\multicolumn{5}{l}{\textit{Panel C: Weight-estimation error $\|\widehat{\bm w}-\widetilde{\bm w}_\sigma\|_1$}} \\
Network & 20  & \textbf{.253 (.002)} & .120 (.001) & .053 (.001) \\
              & 100 & .121 (.001) & \textbf{.055 (.001)} & .024 ($<$.001) \\
              & 500 & .054 (.001) & .025 ($<$.001) & \textbf{.011 ($<$.001)} \\
SPD     & 20  & \textbf{.301 (.002)} & .174 (.002) & .086 (.001) \\
              & 100 & .140 (.001) & \textbf{.081 (.001)} & .039 ($<$.001) \\
              & 500 & .062 ($<$.001) & .036 ($<$.001) & \textbf{.018 ($<$.001)} \\
\multicolumn{5}{l}{\textit{Panel D: Oracle-weight gap $\|\widetilde{\bm w}_\sigma-\bm w^*\|_1$}} \\
Network & 20  & \textbf{.194 (.002)} & .063 (.001) & .015 ($<$.001) \\
              & 100 & .193 (.002) & \textbf{.062 (.001)} & .014 ($<$.001) \\
              & 500 & .194 (.002) & .062 (.001) & \textbf{.015 ($<$.001)} \\
SPD     & 20  & \textbf{.315 (.003)} & .115 (.001) & .029 ($<$.001) \\
              & 100 & .320 (.003) & \textbf{.118 (.001)} & .030 ($<$.001) \\
              & 500 & .316 (.003) & .115 (.001) & \textbf{.029 ($<$.001)} \\
\hline
\end{tabular}
\caption{Network and SPD estimation results for $T_0\in\{20,100,500\}$ and $\sigma\in\{20^{-1/2},100^{-1/2},500^{-1/2}\}$, based on 500 replications. The post-treatment prediction gap compares the post-treatment predictions based on $\widehat{\bm w}$ and $\widetilde{\bm w}_\sigma$, while the realized post-treatment error compares the untreated post-treatment outcome with the prediction based on $\widehat{\bm w}$. The remaining panels report the $\ell_1$ weight-estimation error and oracle-weight gap. Entries are Monte Carlo means with Monte Carlo standard errors in parentheses. Bold entries satisfy $\sigma=T_0^{-1/2}$.}
\label{tab:sim-perturbation-sensitivity}
\end{table}

At each fixed $\sigma$, the post-treatment prediction gap and weight-estimation error decrease as $T_0$ increases, whereas the realized post-treatment error and oracle-weight gap remain stable. Along the diagonal $\sigma=T_0^{-1/2}$, the realized post-treatment errors retain an approximately $T_0^{-1/2}$ pattern: they are $.846$, $.373$, and $.164$ for networks and $.861$, $.386$, and $.173$ for SPD matrices. The post-treatment prediction gaps decrease more rapidly along the same diagonal, from $.161$ to $.031$ and $.006$ for networks and from $.107$ to $.022$ and $.005$ for SPD matrices, an empirical pattern close to $T_0^{-1}$ in these designs. Thus, in these designs, smaller perturbations and longer pre-treatment periods jointly reduce the post-treatment prediction gap and realized post-treatment error, while the newly realized post-treatment perturbation remains the leading component of the latter.

\begin{figure}[tb]
\single
\centering
\includegraphics[width=.78\linewidth]{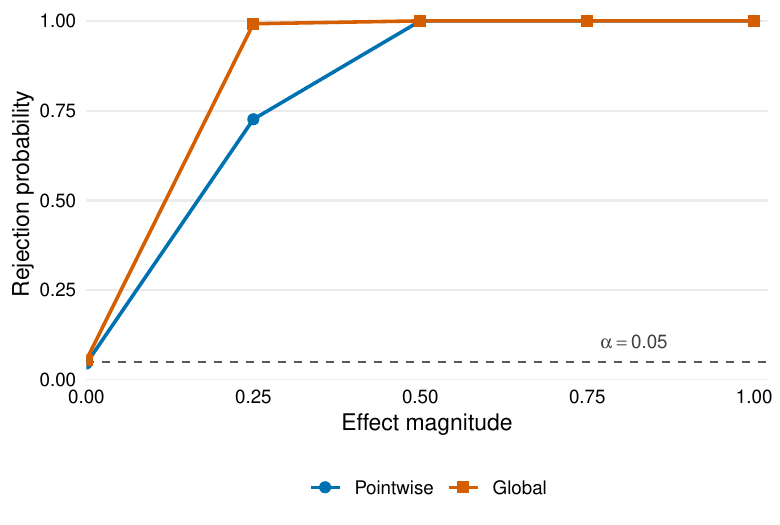}
\caption{Empirical rejection probabilities for the pointwise and global tests in the SPD design over 500 replications. The dashed line marks the nominal level $\alpha=.05$; values at zero effect report empirical size.}
\label{fig:sim-spd-inference}
\end{figure}

The SPD inference experiment uses the same training-calibration split, post-treatment length, and effect grid as the network experiment and sets the global block score to $r=2$. In log coordinates, the imposed treatment satisfies
\[
Y_{1,t}(I)=\exp\!\left\{\log Y_{1,t}(N)+aI_{10}/\sqrt{10}\right\},
\]
so its Log-Euclidean magnitude is exactly $a$. Under the null, pointwise counterfactual coverage is $.954$, and pointwise and global rejection probabilities are $.046$ and $.056$, respectively. At $a=.25$, pointwise and global power are $.726$ and $.992$, and both equal one for $a\geq.50$. Figure~\ref{fig:sim-spd-inference} displays the rejection probabilities.

For the employment comparison, fitting conventional SCM separately to each of the three shares produces the raw post-treatment predictions summarized in Table~\ref{tab:componentwise-employment}. The separate fits do not impose a common donor-weight vector and therefore do not preserve the unit-sum constraint. Closing the three predictions is an additional post-processing step that changes the componentwise estimates; the maximum change to a single component is $.003642$ over the held-out pre-treatment period and $.001996$ after treatment.

\begin{table}[H]
\single
\centering
\begin{tabular}{lcccc}
\hline
Year & Primary & Secondary & Tertiary & Sum \\
\hline
2011 & .055462 & .237262 & .708622 & 1.001346 \\
2012 & .053831 & .237325 & .711657 & 1.002813 \\
2013 & .052093 & .229994 & .716648 & .998735 \\
\hline
\end{tabular}
\caption{Raw predictions from three separately fitted componentwise SCMs in the employment application.}
\label{tab:componentwise-employment}
\end{table}

Table~\ref{tab:application-inference} gives the complete pointwise inference results for both applications. The $90\%$ split-conformal intervals use 11 calibration periods for employment and 10 for fertility. At the $95\%$ level, the split-conformal quantile is $+\infty$ in both applications because the calibration samples are too short to attain that rank.

\begin{table}[H]
\single
\centering
\begin{tabular}{llccc}
\hline
Application & Period & $\widehat\delta_t$ & $90\%$ interval & $p$-value \\
\hline
Employment & 2011 & .0421 & $[.0066,.0776]$ & $1/12$ \\
 & 2012 & .0561 & $[.0206,.0916]$ & $1/12$ \\
 & 2013 & .0646 & $[.0291,.1001]$ & $1/12$ \\
Fertility & 1972 & .0813 & $[.0207,.1419]$ & $1/11$ \\
 & 1973 & .1279 & $[.0673,.1885]$ & $1/11$ \\
 & 1974 & .1544 & $[.0938,.2150]$ & $1/11$ \\
 & 1975 & .1332 & $[.0726,.1938]$ & $1/11$ \\
\hline
\end{tabular}
\caption{Pointwise effect-magnitude estimates, split-conformal intervals, and conformal $p$-values for the applications.}
\label{tab:application-inference}
\end{table}

\end{document}